\documentclass[reprint, superscriptaddress, prb]{revtex4-1}
\usepackage{lineno}
\usepackage{graphicx}
\usepackage{float}
\usepackage{amssymb}
\usepackage{mathrsfs}
\usepackage[T1]{fontenc}
\usepackage{hyperref}
\usepackage[british]{babel}

\begin{document}
\title{Deterministic quantum teleportation between distant superconducting chips}
\author{Jiawei Qiu}
\thanks{These authors contributed equally: Jiawei Qiu, Yang Liu, Jingjing Niu, Ling Hu}
\affiliation{Shenzhen Institute for Quantum Science and Engineering, Southern University of Science and Technology, Shenzhen 518055, China}
\affiliation{International Quantum Academy, Shenzhen 518048, China}
\affiliation{Guangdong Provincial Key Laboratory of Quantum Science and Engineering, Southern University of Science and Technology, Shenzhen 518055, China}
\author{Yang Liu}
\thanks{These authors contributed equally: Jiawei Qiu, Yang Liu,  Jingjing Niu, Ling Hu}
\affiliation{Shenzhen Institute for Quantum Science and Engineering, Southern University of Science and Technology, Shenzhen 518055, China}
\affiliation{International Quantum Academy, Shenzhen 518048, China}
\affiliation{Guangdong Provincial Key Laboratory of Quantum Science and Engineering, Southern University of Science and Technology, Shenzhen 518055, China}
\author{Jingjing Niu}
\thanks{These authors contributed equally: Jiawei Qiu, Yang Liu,  Jingjing Niu, Ling Hu}
\affiliation{Shenzhen Institute for Quantum Science and Engineering, Southern University of Science and Technology, Shenzhen 518055, China}
\affiliation{International Quantum Academy, Shenzhen 518048, China}
\affiliation{Guangdong Provincial Key Laboratory of Quantum Science and Engineering, Southern University of Science and Technology, Shenzhen 518055, China}
\author{Ling Hu}
\thanks{These authors contributed equally: Jiawei Qiu, Yang Liu,  Jingjing Niu, Ling Hu}
\affiliation{Shenzhen Institute for Quantum Science and Engineering, Southern University of Science and Technology, Shenzhen 518055, China}
\affiliation{International Quantum Academy, Shenzhen 518048, China}
\affiliation{Guangdong Provincial Key Laboratory of Quantum Science and Engineering, Southern University of Science and Technology, Shenzhen 518055, China}
\author{Yukai Wu}
\affiliation{Center for Quantum Information, Institute for Interdisciplinary Information Sciences, Tsinghua University, Beijing 100084, PR China}
\author{Libo Zhang}
\affiliation{Shenzhen Institute for Quantum Science and Engineering, Southern University of Science and Technology, Shenzhen 518055, China}
\affiliation{International Quantum Academy, Shenzhen 518048, China}
\affiliation{Guangdong Provincial Key Laboratory of Quantum Science and Engineering, Southern University of Science and Technology, Shenzhen 518055, China}
\author{Wenhui Huang}
\affiliation{Shenzhen Institute for Quantum Science and Engineering, Southern University of Science and Technology, Shenzhen 518055, China}
\affiliation{International Quantum Academy, Shenzhen 518048, China}
\affiliation{Guangdong Provincial Key Laboratory of Quantum Science and Engineering, Southern University of Science and Technology, Shenzhen 518055, China}
\author{Yuanzhen Chen}
\affiliation{Shenzhen Institute for Quantum Science and Engineering, Southern University of Science and Technology, Shenzhen 518055, China}
\affiliation{International Quantum Academy, Shenzhen 518048, China}
\affiliation{Guangdong Provincial Key Laboratory of Quantum Science and Engineering, Southern University of Science and Technology, Shenzhen 518055, China}
\affiliation{Department of Physics, Southern University of Science and Technology, Shenzhen 518055, China}
\author{Jian Li}
\affiliation{Shenzhen Institute for Quantum Science and Engineering, Southern University of Science and Technology, Shenzhen 518055, China}
\affiliation{International Quantum Academy, Shenzhen 518048, China}
\affiliation{Guangdong Provincial Key Laboratory of Quantum Science and Engineering, Southern University of Science and Technology, Shenzhen 518055, China}
\author{Song Liu}
\email{lius3@sustech.edu.cn}
\affiliation{Shenzhen Institute for Quantum Science and Engineering, Southern University of Science and Technology, Shenzhen 518055, China}
\affiliation{International Quantum Academy, Shenzhen 518048, China}
\affiliation{Guangdong Provincial Key Laboratory of Quantum Science and Engineering, Southern University of Science and Technology, Shenzhen 518055, China}
\author{Youpeng Zhong}
\email{zhongyp@sustech.edu.cn}
\affiliation{Shenzhen Institute for Quantum Science and Engineering, Southern University of Science and Technology, Shenzhen 518055, China}
\affiliation{International Quantum Academy, Shenzhen 518048, China}
\affiliation{Guangdong Provincial Key Laboratory of Quantum Science and Engineering, Southern University of Science and Technology, Shenzhen 518055, China}
\author{Luming Duan}
\email{lmduan@tsinghua.edu.cn}
\affiliation{Center for Quantum Information, Institute for Interdisciplinary Information Sciences, Tsinghua University, Beijing 100084, PR China}
\author{Dapeng Yu}
\email{yudp@sustech.edu.cn}
\affiliation{Shenzhen Institute for Quantum Science and Engineering, Southern University of Science and Technology, Shenzhen 518055, China}
\affiliation{International Quantum Academy, Shenzhen 518048, China}
\affiliation{Guangdong Provincial Key Laboratory of Quantum Science and Engineering, Southern University of Science and Technology, Shenzhen 518055, China}
\affiliation{Department of Physics, Southern University of Science and Technology, Shenzhen 518055, China}
\maketitle

\textbf{
Quantum teleportation~\cite{Bennett1993} is of both fundamental interest and great practical importance in quantum information science.
To date, quantum teleportation has been implemented in various physical systems~\cite{Pirandola2015}, among which superconducting qubits are of particular practical significance as they emerge as a leading system to realize large-scale quantum computation~\cite{Arute2019,Wu2021}.
Nevertheless, the number of superconducting qubits on the same chip is severely limited by the available chip size, the cooling power, and the wiring complexity.
Realization of quantum teleportation and remote computation over qubits on distant superconducting chips is a key quantum communication technology to scaling up the system through a distributed quantum computational network~\cite{Gottesman1999,Eisert2000,Jiang2007,Kimble2008,Monroe2014}.
However, this goal has not been realized yet in experiments due to the technical challenge of making a quantum interconnect between distant superconducting chips and the inefficient transfer of flying microwave photons over the lossy interconnects~\cite{Kurpiers2018,Axline2018,Campagne2018,Magnard2020}.
Here we demonstrate deterministic teleportation of quantum states and entangling gates between distant superconducting chips connected by a 64-meter-long cable bus featuring an ultralow loss of 0.32~dB/km at cryogenic temperatures, where high fidelity remote entanglement is generated via flying microwave photons utilizing time-reversal-symmetry~\cite{Cirac1997,Korotkov2011}. 
Apart from the fundamental interest of teleporting macroscopic superconducting qubits over a long distance, our work lays a foundation to realization of large-scale superconducting quantum computation through a distributed computational network~\cite{Gottesman1999,Eisert2000,Jiang2007,Kimble2008,Monroe2014}.
}

Exploiting the physical resource of entanglement to transfer unknown quantum states over long distances without physically moving the object itself, quantum teleportation~\cite{Bennett1993} is one of the most profound protocols in quantum information science.
Apart from its fundamental interest, teleportation also plays a crucial role in long-distance quantum communication~\cite{Duan2001} and distributed computational networks~\cite{Gottesman1999,Eisert2000,Jiang2007,Kimble2008,Monroe2014}.
Although the teleportation of large objects remains a fantasy, experimental realization of quantum teleportation has been implemented with optical photons over meter-scale distances in laboratory first~\cite{Bouwmeester1997,Furusawa1998}, then over kilometre-scale distances in free space~\cite{Ma2012}, and more recently over thousand-kilometer-scale distances through ground-to-satellite communication~\cite{Ren2017}.
In atomic systems, fully deterministic quantum teleportation has been realized over micrometre-scale distances with ions in the same trap~\cite{Riebe2004,Barrett2004}, whereas a probabilistic protocol has also been implemented between ions in different traps separated by a meter~\cite{Olmschenk2009}.
In solid-state systems, quantum teleportation has been implemented with spin qubits in diamonds separated by three meters unconditionally~\cite{Pfaff2014}, and more recently between non-neighbouring quantum nodes probabilistically~\cite{Hermans2022}.

\begin{figure*}[t]
\begin{center}
	\includegraphics[width=0.8\textwidth]{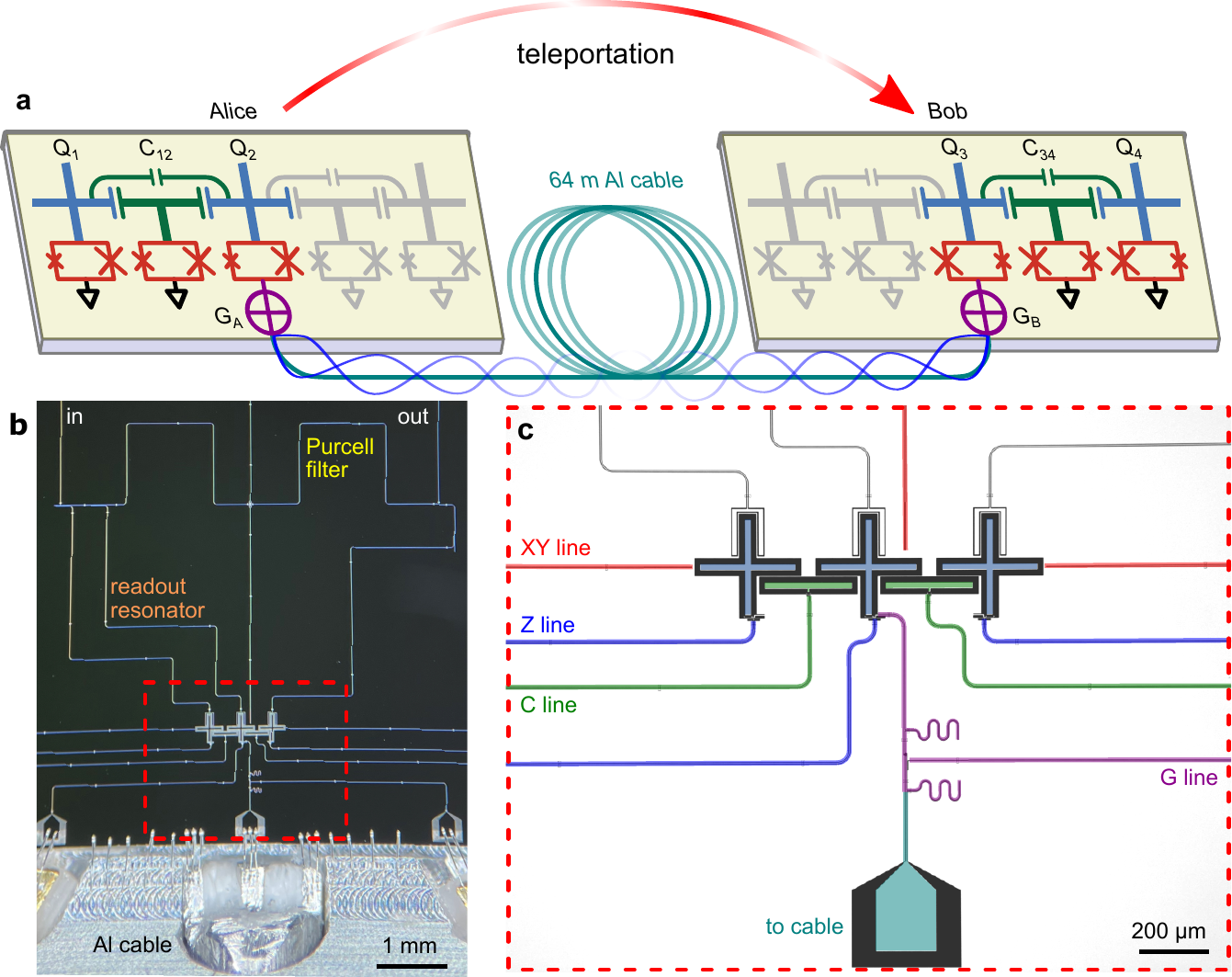}
	\caption{
    \label{fig1}
    {\bf Description of the superconducting quantum network.}
    {\bf a,} Schematic of the quantum network, where two nodes Alice and Bob are connected through a 64-meter-long Al cable. Exploiting the physical resource of remote entanglement between $Q_2$ and $Q_3$, quantum states and entangling gates are teleported from Alice to Bob.
    {\bf b,} Photograph of a quantum processor node.
    {\bf c,} Zoomed in micrograph of {\bf b}, showing the qubits, couplers and control/readout circuitries. Three X-shaped transmon qubits (shaded in light blue) are coupled together via T-shaped transmon tunable couplers (shaded in light green); the central qubit is galvanically connected to a gmon tunable coupler (shaded in light purple) for cable connection.
    }
\end{center}
\end{figure*}

Marked by the demonstration of quantum supremacy~\cite{Arute2019} or quantum advantage~\cite{Wu2021}, superconducting circuits have seen remarkable progresses towards scalable quantum computation in the past few years.
However, along with the scaling comes formidable emerging technical challenges such as the available chip size, the cooling power, and the wiring complexity.
Distributed quantum computational networks based on superconducting qubits~\cite{Kurpiers2018,Axline2018,Campagne2018,Magnard2020,Zhong2021} promise further scalability beyond single chip or even single cryostat~\cite{Nickerson2014,Bravyi2022}, 
but owing to the inherent isolation between spatially separate modules, multi-qubit operations in the modular architecture cannot be implemented using direct interactions.
Quantum teleportation can transfer unknown quantum states and implement entangling operations between two qubits housed in spatially separate modules without the need for any direct interaction, making it an essential tool for universal quantum computation in the modular architecture~\cite{Gottesman1999,Eisert2000,Jiang2007}.
Moreover, it is also of great fundamental interest to teleport the states of macroscopic superconducting qubits visible to the naked eye over long distances.
Unfortunately, despite the technical utility and scientific appeal of such a capability, quantum teleportation with superconducting qubits has been limited on the same chip or module up to a distance of a few millimeters thus far~\cite{Steffen2013,Chou2018}.
Coherent optical links for superconducting qubits can herald entanglement between distant superconducting chips housed in separate cryostats, but it still suffers from massive conversion loss in microwave-to-optical transduction at single photon levels~\cite{Mirhosseini2020}.
Flying microwave photons propagating through cryogenic electrical cables/waveguides provide a direct, and so far the only viable means for connecting distant superconducting chips.
Quantum teleportation of propagating coherent microwave states has been recently demonstrated with two-mode squeezing and analog feedforward over a macroscopic distance~\cite{Fedorov2021}.
However, the fidelities of quantum state transfer and remote entanglement using flying microwave photons have been confined to $\sim80\%$, hindered by the lossy interconnects~\cite{Kurpiers2018,Axline2018,Campagne2018,Magnard2020}. With the development of low-loss interconnects based on superconducting aluminum coaxial cables and on-chip impedance transformers, 99\% fidelity has been achieved using standing modes over short distances recently~\cite{Niu2023}.

Using remote entanglement to implement non-local quantum teleportation between distant superconducting chips has several demanding requirements which have not been simultaneously satisfied to date.
Here, we overcome this long-standing challenge by combining several technical advancements together and report the successful implementation of deterministic quantum teleportation between distant superconducting chips separated by a 64-meter microwave cable. First, we connect two quantum chips using a long superconducting cable featuring an ultralow loss of 0.32~dB/km for microwave photons at cryogenic temperatures, comparable to the typical performance of 0.2 dB/km for optical fibers. Photons lose only 0.5\% of their energies in each transit through the cable, thus resolving the interconnect bottleneck. Second, by using gmon tunable couplers~\cite{Chen2014} to shape and catch flying microwave photons in a time-reversal-symmetric manner~\cite{Cirac1997,Korotkov2011}, we generate a remote Einstein-Podolsky-Rosen (EPR) pair with $94.2\pm 0.6\%$ fidelity. To our knowledge, this is a record-high fidelity for remote EPR pairs generated using flying microwave photons.
Note with similar low loss interconnects, 99\% fidelity has been recently achieved using standing modes over 0.25~m distance~\cite{Niu2023}.
Finally, we apply several noise-reduction techniques to the superconducting chips, including integrated Purcell filters for high fidelity single-shot qubit readout~\cite{Jeffrey2014}, and asymmetric qubit junctions for long coherence time~\cite{Hutchings2017}. Microwave electronics based on field programmable gate array (FPGA) with branching and real-time feed-back/forward are used for triggering the remote qubit state rotation pulses. All these advanced techniques combined together allow our realization of the deterministic teleportation of quantum states over different chips with an average process fidelity of $78.3\pm 0.8\%$, significantly above the 1/2 classical threshold, unequivocally demonstrating its quantum nature.
We further implement the deterministic teleportation of a control-NOT (CNOT) gate, with an average process fidelity of $70.2\pm 0.6\%$.
Apart from the fundamental interest of teleporting macroscopic superconducting qubits over a large distance, our work lays a foundation to realization of large-scale superconducting quantum computation through a distributed computational network~\cite{Gottesman1999,Eisert2000,Jiang2007}.

\begin{figure*}[t]
\begin{center}
	\includegraphics[width=0.85\textwidth]{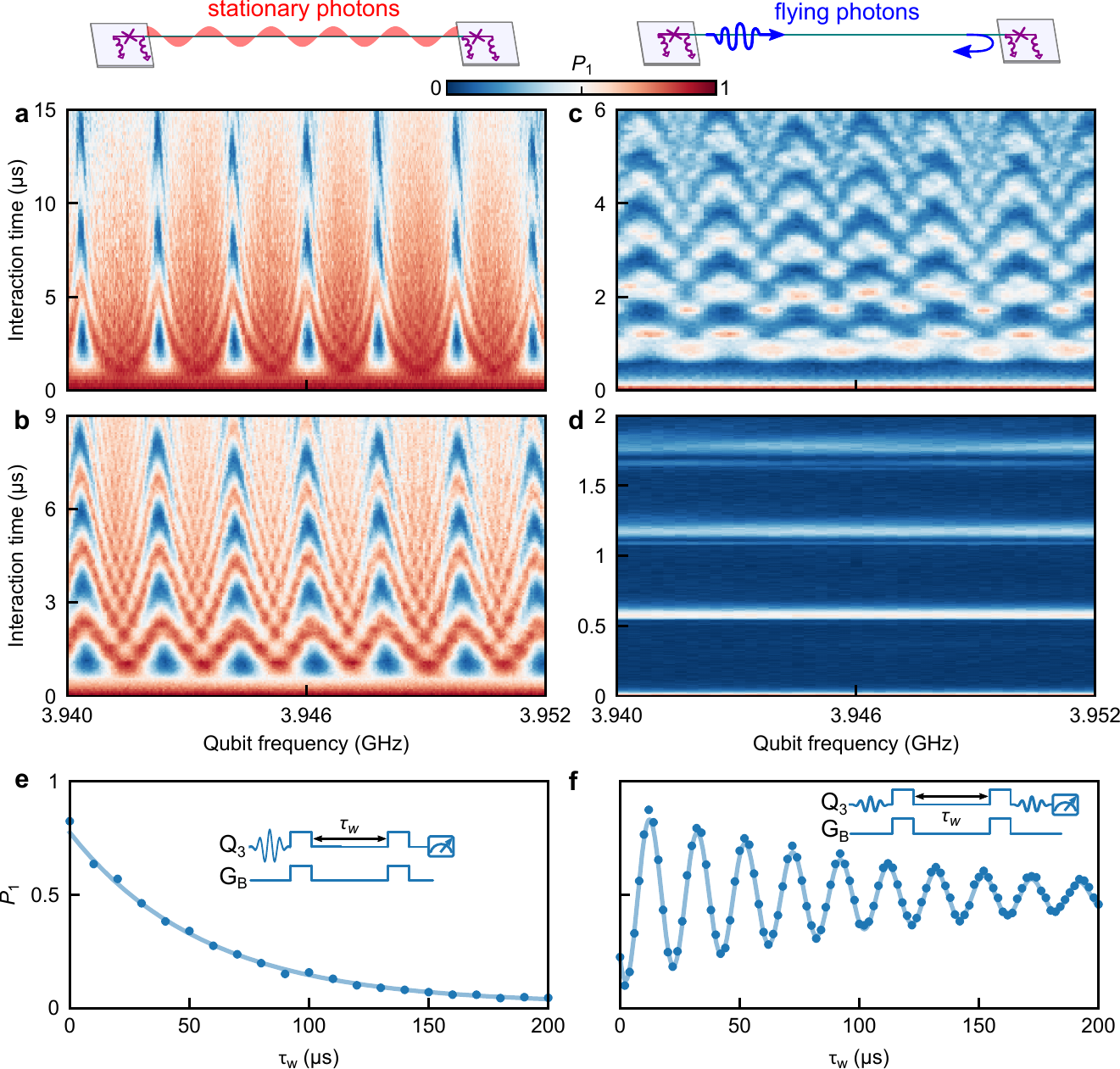}
	\caption{
    \label{fig2}
    {\bf Transition from stationary to flying microwave photons.}
    {\bf a}--{\bf d}, Interaction between $Q_3$ and the long cable at different coupling strengths. At weak multimode coupling ({\bf a}--{\bf b}), the qubit can independently address each standing wave mode, where the photons are stationary in the cable as illustrated on top. At strong multimode coupling ({\bf c}--{\bf d}), adjacent standing wave modes merge together and the chevron pattern of vacuum Rabi oscillations fades, instead a stripe pattern emerges, featuring a ping-pong dynamics where the emitted flying microwave photons are bounced back and forth in the cable, as illustrated on top.
    {\bf e}, {\bf f} Measuring the lifetime $T_{1}^{m}$ and dephasing time $T_{2}^m$ of a standing wave mode. Inset: control pulse sequence. Fitting the data gives $T_{1}^m=56.2$~$\mu$s, $T_2^m=106.8$~$\mu$s $\approx 2 T_1^m$, suggesting a linear loss of 0.32 dB/km and negligible dephasing noise in the cable.
    }
\end{center}
\end{figure*}

Our quantum network consists of two nodes Alice and Bob, see Fig.~\ref{fig1}, where each node has three X-shaped transmon qubits~\cite{Barends2013} coupled by T-shaped transmon couplers~\cite{Yan2018}. The central qubit in each node is galvanically connected to a gmon tunable coupler~\cite{Chen2014} $G_n$ ($n=A$, $B$) for cable connection, where $G_n$ is placed close to the bonding pad.
In this experiment, two qubits $Q_1$ and $Q_2$ and their coupler $C_{12}$ in Alice, $Q_3$ and $Q_4$ and their coupler $C_{34}$ in Bob, and the two gmon couplers $G_A$ and $G_B$ are used, as shown in Fig.~\ref{fig1}{\bf a}.
The 64-meter-long cable is made of pure aluminum (Al), the same material as the superconducting circuits.
It is joined by four sections using wirebond connection, where each section is 16 meters long limited by the manufacturing process.
More details about the device and experimental setup can be found in the Supplementary Information.

\begin{figure*}[t]
\begin{center}
	\includegraphics[width=0.85\textwidth]{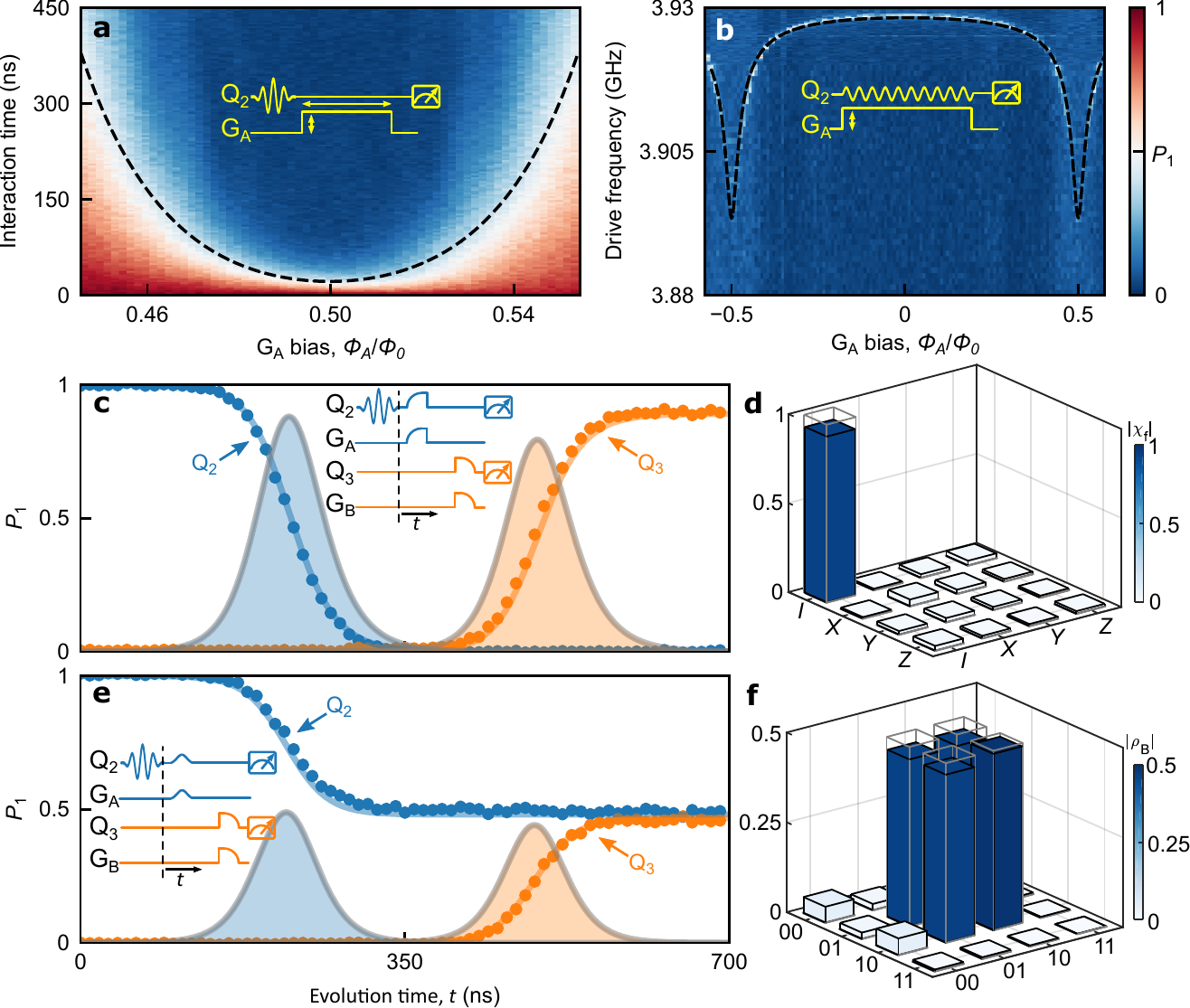}
	\caption{
    \label{fig3}
    {\bf Efficient pitch-and-catch of flying microwave photons.}
    {\bf a,} Controlling the $Q_2$ photon emission rate $\kappa_A$ with the $G_A$ flux bias $\Phi_A$. The dashed line is $1/\kappa_A$ obtained from fitting.
    {\bf b,} $Q_2$ spectroscopy, revealing its frequency shift induced by the change of $G_A$ junction inductance.
    {\bf c,} Time-reversal-symmetric transfer of a flying photon from $Q_2$ to $Q_3$, with a transfer efficiency $\eta = 90.4 \pm 0.8\%$. Filled shapes: emitted (light blue) and received (light orange) flying photon envelopes.
    {\bf d,} Process matrix $\chi_f$ of the state transfer in {\bf c}, with a fidelity of $\mathcal{F}_{f}^{p}=93.2 \pm 0.5\%$. The horizontal axes are the Pauli operators $I$, $X$, $Y$ and $Z$.
    {\bf e,} Remote entanglement generation by emitting half a photon from $Q_2$ to $Q_3$. Filled shapes: flying photon envelopes.
    {\bf f,} Density matrix of the Bell state generated in {\bf e}, with a state fidelity of $\mathcal{F}_B =94.2\pm 0.6\%$.
    Insets to {\bf a}, {\bf b}, {\bf c} and {\bf e} are control pulse sequences, where the qubit bias pulses are for frequency compensation; solid lines in {\bf c}, {\bf e} are simulations; solid bars and grey frames in {\bf d}, {\bf f} are measured and ideal values respectively. Data in {\bf c}--{\bf f} are corrected for readout errors.
    }
\end{center}
\end{figure*}

We first explore the quantum dynamics between the qubits and the long cable.
The cable itself naturally supports a sequence of standing wave modes evenly-spaced in spectrum, with a free spectral range $\omega_{\rm FSR}/2\pi \approx 1.89$~MHz determined by the cable length, corresponding to a single transit time of $\tau_{st} = \pi/\omega_{\rm FSR} = 265$~ns.
For convenience, we use $Q_3$ in Bob to characterize the cable, varying $G_B$ coupler flux bias $\Phi_B$ while keeping $G_A$ off.
Denote the Jaynes-Cummings coupling strength between $Q_3$ and the $m$-th standing wave mode at $\sim$3.94~GHz as $g^m_B$ ($m\sim 2085$), at weak multimode coupling regime, i.e., $g^m_B \ll \omega_{\rm FSR}$, the qubit can independently address each mode, where the photons are stationary in the cable (see Fig.~\ref{fig2}{\bf a}, {\bf b} where $g^m_B/2\pi=$ 0.08~MHz, 0.25~MHz respectively).
At strong multimode coupling regime, where $g^m_B\sim \omega_{\rm FSR}$, adjacent standing wave modes merge together and the chevron pattern of vacuum Rabi oscillations fades out, instead a stripe pattern emerges (see Fig.~\ref{fig2}{\bf c}, {\bf d} where $g^m_B/2\pi=$ 0.45~MHz, 1.63~MHz respectively), featuring a ping-pong dynamics where the emitted flying photons are bounced back and forth in the cable~\cite{Zhong2019,Bienfait2019}.
In this regime, it is more appropriate to describe the quantum dynamics using input-output theory~\cite{Gardiner1985}, where the qubit-cable coupling is characterized by the photon emission rate $\kappa_B$ (see Supplementary Information for details), which is related to the Jaynes-Cummings coupling strength $g_B^m$ via Fermi's golden rule: $\kappa_B=2\pi (g^m_B)^2/\omega_{\rm FSR}$.
The large electrical delays in photon emission have also been explored with the help of metamaterials~\cite{Ferreira2021,Ferreira2022} or surface-acoustic-wave transducers~\cite{Andersson2019,Bienfait2019}.
We measure the lifetime of the $m=2087$-th mode at 3.9441~GHz by swapping a photon into the mode at weak multimode coupling, waiting for a period $\tau_w$, then swapping out the photon and measuring the qubit excited state $|1\rangle$ population $P_1$ versus $\tau_w$, see Fig.~\ref{fig2}{\bf e}.
Fitting the exponential decay gives a mode lifetime of $T_1^m=56.2$~$\mu$s, corresponding to a quality factor of $1.4 \times10^6$ and a linear loss of $\alpha_c = 8.686$~dB$/(2 v_c T_{1}^m)=0.32$~dB/km, where 8.686~dB is 1 Neper, and $v_c\approx 2.4\times 10^8$~m/s is the speed of light in the cable.
We note that the cable mode lifetime is even longer than that of the qubits in this experiment, highlighting its ultralow loss feature.
Such ultralow-loss quantum interconnects could become the backbone of future quantum networks based on superconducting circuits~\cite{Awschalom2021,Bravyi2022}.
Limited by space, the long cable is coiled up inside the dilution fridge (see Supplementary Information), as such, low frequency fluctuation in the environmental magnetic field could couple with the cable modes via electromagnetic induction and cause dephasing.
We perform Ramsey interference with the $m$-th mode and find that $T_2^m=106.8$~$\mu$s $\approx 2 T_1^m$ (Fig.~\ref{fig2}{\bf f}), suggesting negligible dephasing noise.

\begin{figure*}[t]
\begin{center}
	\includegraphics[width=0.85\textwidth]{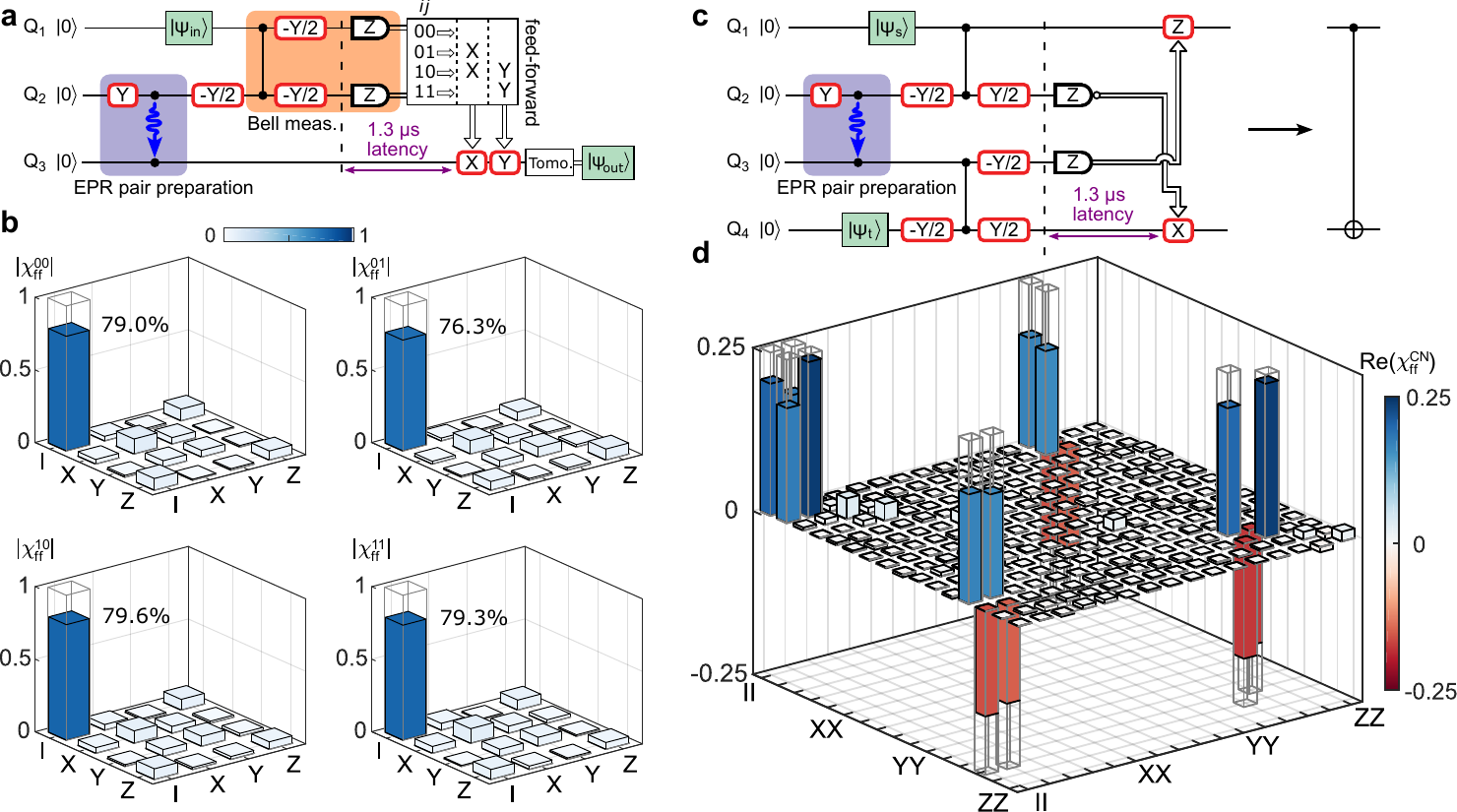}
	\caption{
    \label{fig4}
    {\bf Deterministic quantum teleportation with feed-forward.}
    {\bf a}, {\bf b}, The protocol and process matrix  $\chi_{{\rm ff}}^{ij}$ for deterministic quantum state teleportation from $Q_1$ to $Q_3$ with feed-forward. The average process fidelity is $78.3\pm 0.8\%$.
    {\bf c}, {\bf d}, The protocol and process matrix  $\chi_{{\rm ff}}^{CN}$ for deterministic teleportation of the CNOT gate, with a process fidelity of $70.2\pm 0.6\%$.
    The solid bars and grey frames in {\bf b}, {\bf d} are the measured and ideal values respectively.
    }
\end{center}
\end{figure*}

Although single photon detection has been well developed at optical frequencies, capturing flying microwave photons with high efficiency remains a long-standing challenge as their photon energies are four or five orders of magnitude lower~\cite{Besse2018,Lescanne2020}.
It is proposed that by dynamically tuning the sender coupling to symmetrize the emitted photon envelope, and then tuning the receiver coupling reversely, 100\% transfer efficiency could be achieved because of time-reversal-symmetry~\cite{Cirac1997,Korotkov2011}.
Using microwave parametric driving processes for flying photon shaping and catching, recent demonstrations have achieved $\sim 80$\% quantum state transfer fidelities between distant superconducting chips connected by channels up to five meters long~\cite{Kurpiers2018,Axline2018,Campagne2018,Magnard2020}.
Alternatively, on-chip demonstrations using flux-controlled gmon tunable couplers~\cite{Chen2014} provide larger coupling tuning range and have shown great potential in shaping and catching flying microwave photons~\cite{Zhong2019} as well as phonons~\cite{Bienfait2019}.
Now we demonstrate high fidelity pitch-and-catch of flying microwave photons through the 64~m ultralow-loss cable.
In Fig.~\ref{fig3}{\bf a}, we characterize the photon emission rate by varying the $G_A$ flux bias $\Phi_A$.
At $\Phi_A=\Phi_0/2$ where $\Phi_0=2.067\times 10^{-15}$~Wb is a flux quantum, a maximum emission rate $\kappa_A^{\rm max}=1/22$~ns$^{-1}$ is achieved.
$G_B$ is also characterized in a similar way, with $\kappa_B^{\rm max}=1/18$~ns$^{-1}$, slightly larger than $G_A$, see Supplementary Information.
Note that the qubit frequency can be shifted by tens of MHz as the coupler junction inductance is changed, see Fig.~\ref{fig3}{\bf b}.
This frequency shift has a strong impact on photon transfer efficiency~\cite{Korotkov2011,Sete2015} and must be counteracted precisely with the qubit Z bias~\cite{Zhong2019}.
Prior to the quantum state transfer process, an active cooling procedure is performed to reduce the thermal photons in the cable and the qubits (see Supplementary Information for details).
Following Refs.~\onlinecite{Kurpiers2018,Bienfait2019}, we dynamically tune the photon emission rate as
\begin{equation}\label{kappa}
  \kappa_A (t) = \kappa_c \frac{e^{\kappa_c t}}{1+e^{\kappa_c t}},
\end{equation}
emitting a flying photon with a symmetric envelope $\propto \rm{sech}(\kappa_c t/2)$, here $\kappa_c=\kappa_A^{\rm max}=1/22$~ns$^{-1}$ is the characteristic emission rate during the photon transfer process.
Tuning the $G_B$ coupling rate reversely, i.e., $\kappa_B(t) = \kappa_A(\tau_{st}-t)$, we transfer a flying microwave photon from $Q_2$ to $Q_3$ with efficiency $\eta = 90.4\pm 0.8\%$, see Fig.~\ref{fig3}{\bf c}.
A quantitative analysis of the transfer inefficiency is provided in the Supplementary Information.
We perform quantum process tomography~\cite{Zhong2021} to characterize this process, yielding a process matrix $\chi_f$ shown in Fig.~\ref{fig3}{\bf d} with a fidelity $\mathcal{F}_{f}^{p}=\textrm{Tr}(\chi_{f}\cdot \mathcal{I})=93.2\pm 0.5\%$, where $\mathcal{I}$ is the identity operation.
Note all uncertainties reported here represent the standard deviation of repeated measurements.
To generate remote entanglement between $Q_2$ and $Q_3$, we emit half a photon from $Q_2$ to the cable (see Supplementary Information for details), and capture the photon with $Q_3$, see Fig.~\ref{fig3}{\bf e}.
Ideally, a Bell state $|\psi_B\rangle=(|01\rangle+|10\rangle)/\sqrt{2}$ is generated between $Q_2$ and $Q_3$, here $|0\rangle$ is the qubit ground state.
The density matrix $\rho_B$ of the Bell state is measured with quantum state tomography~\cite{Zhong2021} and shown in Fig.~\ref{fig3}{\bf f}, with a state fidelity $\mathcal{F}_B=\langle \psi_B|\rho_B|\psi_B\rangle=94.2\pm 0.6\%$.
To our knowledge, this is a record-high fidelity for EPR pairs generated using flying microwave photons, even compared with monolithic results~\cite{Zhong2019}.

A standard quantum teleportation protocol consists of three steps:
(i) the creation of an EPR pair shared between Alice and Bob, which has been successfully demonstrated in Fig.~\ref{fig3}{\bf e};
(ii) high fidelity entangling gate and single-shot Bell measurement in Alice;
(iii) feed-forward of the (classical) measurement information in step (ii) to perform the remote qubit state rotations in Bob.
In this experiment, with the help of on-chip Purcell filters~\cite{Jeffrey2014}, the qubits are strongly coupled with their readout resonators without sacrificing their coherence, allowing for high fidelity, single-shot dispersive readout (see Supplementary Information for details), thus meeting the requirement of step (ii).
Experimentally, the feed-forward step incurs a long latency of microsecond level, a considerable amount of time compared to the coherence of superconducting qubits. By sacrificing the deterministic feature, one could avoid this latency and instead post-select the $Q_3$ outcome upon the classical measurement results in step (ii), see Supplementary Information.
Using FPGA-based microwave electronics, we further implement the deterministic quantum teleportation with feed-forward, see the protocol schematic in Fig.~\ref{fig4}{\bf a}.
Defining the state discrimination thresholds for the joint readout of $Q_1$ and $Q_2$ in the FPGA, we feed-forward (ff) the digital measurement outcome $ij$ (00, 01, 10 or 11) to trigger the remote state rotation pulses $X$ or $Y$ for $Q_3$.
The latency of the feed-forward step is as long as 1.3 $\mu$s including the 0.7~$\mu$s dispersive readout time of $Q_1$ and $Q_2$, maintaining the coherence of $Q_3$ during this process is crucial for completing the teleportation successfully.
With this end, the qubits are fabricated with asymmetric Josephson junctions~\cite{Hutchings2017} and operated near their minimal frequency sweet spots where they are insensitive to low frequency dephasing noise.
In particular, $Q_3$ is operated right at its frequency sweet spot with a lifetime of 33.6~$\mu$s and a pure dephasing time of 62.6~$\mu$s, sufficiently long compared to the feed-forward latency.
The teleportation process matrices $\chi_{{\rm ff}}^{ij}$ are shown in Fig.~\ref{fig4}{\bf b}, where $\chi_{{\rm ff}}^{ij}$ has a process fidelity of $79.0\pm 1.2\%$~(00), $76.3\pm 1.3\%$~(01), $79.6\pm 1.3\%$~(10) and $79.3\pm 1.2\%$~(11) respectively. The overall deterministic teleportation process fidelity is $78.3\pm0.8\%$, significantly above the classical threshold of 1/2, unequivocally revealing the quantum nature of this process.

In a modular architecture where separate quantum nodes are connected together via communication channels, the teleportation of an
entangling quantum gate is an essential tool for universal quantum computation\cite{Gottesman1999,Eisert2000,Jiang2007}.
The deterministic teleportation of a CNOT gate have been recently demonstrated using superconducting qubits housed within the same physical module~\cite{Chou2018}, ions in the same trap~\cite{Wan2019}, and distant atoms entangled through a 60 meters long optical fiber~\cite{Daiss2021}.
Here, we implement the deterministic teleportation of a non-local CNOT gate between two distant superconducting chips utilizing remote entanglement, see the protocol schematic in Fig.~\ref{fig4}{\bf c}.
Following the remote entanglement of $Q_2$-$Q_3$, local CNOT gates are applied to the $Q_1$-$Q_2$ and $Q_3$-$Q_4$ qubit pairs.
Before the simultaneous single-shot measurement on $Q_2$-$Q_3$ pair, a $-Y/2$ rotation is applied to $Q_3$ to rotate its X axis to Z axis.
Defining the state discrimination thresholds for the readout of $Q_2$ and $Q_3$ in the FPGA, we feed-forward the measurement outcome 1 of $Q_3$ to trigger the remote state rotation pulse $Z$ for $Q_1$, and the measurement outcome 0 of $Q_2$ to trigger the remote state rotation pulse $X$ for $Q_4$.
This procedure effects a CNOT operation between $Q_1$ in Alice and $Q_4$ in Bob, without a direct interaction between them.
The process matrix $\chi_{{\rm ff}}^{CN}$ for the teleported CNOT gate is characterized by quantum process tomography, where the real part $\textrm{Re}(\chi_{{\rm ff}}^{CN})$ is shown in Fig.~\ref{fig4}{\bf d} with a process fidelity of $70.2\pm 0.6\%$ (see Supplementary Information for the imaginary part).


The advance from teleportation on a single chip or module~\cite{Steffen2013,Chou2018} to teleportation over distant chips, as demonstrated here, is a key step in building large-scale superconducting quantum computation through a distributed computational network~\cite{Gottesman1999,Eisert2000,Jiang2007,Kimble2008,Monroe2014}.
The teleportation fidelities can be further improved with entanglement-purification protocols~\cite{Jiang2007,Bennett1996}, rapid high-fidelity single-shot
readout techniques~\cite{Walter2017} and continuing improvements in coherence time and circuit design.
With a linear loss comparable to optical fibers, the long cable bus developed here could become the backbone of future quantum networks based on superconducting circuits~\cite{Awschalom2021,Bravyi2022}.
Our progress also opens up a new avenue for waveguide quantum electrodynamics~\cite{Kannan2023} and quantum photonics at microwave frequencies~\cite{Gu2017}.
\clearpage

\bibliographystyle{naturemag}
\bibliography{bibliography}

\clearpage


\subsection*{Methods}
\textbf{Standing mode method versus flying photon method.}
Quantum state transfer and entanglement generation through a standing mode has been routinely used on single chips, using half-wavelength coplanar waveguide resonators as communication buses~\cite{Sillanpaa2007,Majer2007,Ansmann2009,Steffen2013}.
For short-distance communication between different quantum chips within the same fridge, for example through microwave cables shorter than a meter, the free spectral range of the cable is large enough (> 100 MHz) that the standing modes are individually addressable, as such, quantum states can be relayed through a single standing mode~\cite{Burkhart2021,Zhong2021}, or via a ``dark'' mode hybridized by a standing mode and the on-chip elements (qubits or resonators)~\cite{Leung2019,Chang2020}.
With the development of low-loss interconnects based on superconducting aluminum coaxial cables and on-chip impedance transformers, 99\% fidelity has been achieved using standing modes over short distances recently~\cite{Niu2023}.
Although the standing mode method is simple and efficient to implement, it is not physically scalable to long distances.
To date, the longest distance quantum state transfer demonstrated with standing mode is 1 meter~\cite{Leung2019,Zhong2021}.
If the length of the cable is increased, the free spectral range becomes inversely smaller, eventually making standing-mode quantum state transfer impractical.
For long distance communication in a local area network of quantum processors housed in separate dilution fridges, flying photon method is more favorable~\cite{Cirac1997,Korotkov2011,Sete2015}.

Highly efficient quantum senders and receivers are needed for deterministic quantum state transfer between distant quantum nodes using flying photons.
The sender converts stationary qubits into flying photons which can travel and share quantum information between the two distant nodes.
The receiver catches the flying photons and convert them to stationary qubits for further processing.
Despite enormous effort, conversion between stationary qubits and flying photons is still particularly challenging.
Naturally emitted flying photons have an exponentially decaying wave packet, for which a receiver with fixed coupling can have at most 54\% absorption efficiency even if it is identical and perfectly matched to the sender~\cite{Stobinska2009,Wang2011}.
It is proposed that by using tunable couplers to dynamically shape the photon envelopes, nearly perfect state transfer efficiency can be achieved as long as the back-refection at the receiver is cancelled out by destructive interference, where time-symmetric pulse is a convenient choice~\cite{Cirac1997,Korotkov2011,Sete2015}.
Experimentally, the controlled emission of flying microwave photons with shaped temporal envelopes are first explored with sender nodes only~\cite{Yin2013,Srinivasan2014,Pechal2014}.
On the other hand, catching time-reversed microwave coherent state photons with high efficiency have been separately demonstrated using external classical microwave sources~\cite{Wenner2014,Lin2022}.
Combining both techniques together, deterministic quantum state transfer and remote entanglement between distant qubits using flying photons are demonstrated in several recent experiments~\cite{Kurpiers2018,Axline2018,Campagne2018,Magnard2020}.
In these experiments, a circulator is used to interrupt the channel for eliminating back-reflection during the tuneup of the flying photon pitch-and-catch process, where the reflection should disappear once the back-refection at the receiver is cancelled out.
The circulator also opens a window for monitoring the waveform of the flying photons, allowing one to calibrate the pulse shape precisely.
However, as a normal metal component, the circulator introduces significant loss, limiting the transfer fidelities to $\sim$80\% in these experiments.
The large cable delay in this experiment renders unnecessary the use of a circulator, but this poses new challenge on the pulse shape calibration. To conquer this challenge, we have developed a non-invasive pulse shape calibration method using photon number dynamics, see Supplementary Information for details.

\subsection*{Acknowledgements}
We thank Junqiu Liu, Yu He and Yuan Xu for helpful discussions.
This work was supported by the Key-Area Research and Development Program of Guangdong Province (2018B030326001), the National Natural Science Foundation of China (U1801661, 12174178, 11905098, 12204228), the Guangdong Innovative and Entrepreneurial Research Team Program (2016ZT06D348), the Guangdong Provincial Key Laboratory (2019B121203002), the Science, Technology and Innovation Commission of Shenzhen Municipality (KYTDPT20181011104202253, KQTD20210811090049034, K21547502), the Shenzhen-Hong Kong Cooperation Zone for Technology and Innovation (HZQB-KCZYB-2020050), and the NSF of Beijing (Z190012). L.D. and Y.W. acknowledge support from the Tsinghua University Initiative Scientific Research Program and the Ministry of Education of China.

\subsection*{Author contributions}
D.Y. and Y.Z. supervised the project.
L.D. and Y.Z. conceived the idea.
J.Q. designed the devices and wirebonded the cables.
J.Q., Y.L., J.N. and L.H. performed the measurements and analyzed the data.
L.Z. fabricated the devices and designed the sample holder supervised by S.L.
Y.Z. wrote the manuscript.
All authors contributed to discussions and production of the manuscript.

\subsection*{Competing interests}
The authors declare no competing interests.

\subsection*{Data availability}
The data that support the plots within this paper and other findings of this study are available from the corresponding author upon reasonable request.
\end{document}


\title{Supplementary Information for ``Deterministic quantum teleportation between distant superconducting chips''}
\author{Jiawei Qiu}
\thanks{These authors contributed equally: Jiawei Qiu, Yang Liu, Jingjing Niu, Ling Hu}
\affiliation{Shenzhen Institute for Quantum Science and Engineering, Southern University of Science and Technology, Shenzhen 518055, China}
\affiliation{International Quantum Academy, Shenzhen 518048, China}
\affiliation{Guangdong Provincial Key Laboratory of Quantum Science and Engineering, Southern University of Science and Technology, Shenzhen 518055, China}
\author{Yang Liu}
\thanks{These authors contributed equally: Jiawei Qiu, Yang Liu,  Jingjing Niu, Ling Hu}
\affiliation{Shenzhen Institute for Quantum Science and Engineering, Southern University of Science and Technology, Shenzhen 518055, China}
\affiliation{International Quantum Academy, Shenzhen 518048, China}
\affiliation{Guangdong Provincial Key Laboratory of Quantum Science and Engineering, Southern University of Science and Technology, Shenzhen 518055, China}
\author{Jingjing Niu}
\thanks{These authors contributed equally: Jiawei Qiu, Yang Liu,  Jingjing Niu, Ling Hu}
\affiliation{Shenzhen Institute for Quantum Science and Engineering, Southern University of Science and Technology, Shenzhen 518055, China}
\affiliation{International Quantum Academy, Shenzhen 518048, China}
\affiliation{Guangdong Provincial Key Laboratory of Quantum Science and Engineering, Southern University of Science and Technology, Shenzhen 518055, China}
\author{Ling Hu}
\thanks{These authors contributed equally: Jiawei Qiu, Yang Liu,  Jingjing Niu, Ling Hu}
\affiliation{Shenzhen Institute for Quantum Science and Engineering, Southern University of Science and Technology, Shenzhen 518055, China}
\affiliation{International Quantum Academy, Shenzhen 518048, China}
\affiliation{Guangdong Provincial Key Laboratory of Quantum Science and Engineering, Southern University of Science and Technology, Shenzhen 518055, China}
\author{Yukai Wu}
\affiliation{Center for Quantum Information, Institute for Interdisciplinary Information Sciences, Tsinghua University, Beijing 100084, PR China}
\author{Libo Zhang}
\affiliation{Shenzhen Institute for Quantum Science and Engineering, Southern University of Science and Technology, Shenzhen 518055, China}
\affiliation{International Quantum Academy, Shenzhen 518048, China}
\affiliation{Guangdong Provincial Key Laboratory of Quantum Science and Engineering, Southern University of Science and Technology, Shenzhen 518055, China}
\author{Wenhui Huang}
\affiliation{Shenzhen Institute for Quantum Science and Engineering, Southern University of Science and Technology, Shenzhen 518055, China}
\affiliation{International Quantum Academy, Shenzhen 518048, China}
\affiliation{Guangdong Provincial Key Laboratory of Quantum Science and Engineering, Southern University of Science and Technology, Shenzhen 518055, China}
\author{Yuanzhen Chen}
\affiliation{Shenzhen Institute for Quantum Science and Engineering, Southern University of Science and Technology, Shenzhen 518055, China}
\affiliation{International Quantum Academy, Shenzhen 518048, China}
\affiliation{Guangdong Provincial Key Laboratory of Quantum Science and Engineering, Southern University of Science and Technology, Shenzhen 518055, China}
\affiliation{Department of Physics, Southern University of Science and Technology, Shenzhen 518055, China}
\author{Jian Li}
\affiliation{Shenzhen Institute for Quantum Science and Engineering, Southern University of Science and Technology, Shenzhen 518055, China}
\affiliation{International Quantum Academy, Shenzhen 518048, China}
\affiliation{Guangdong Provincial Key Laboratory of Quantum Science and Engineering, Southern University of Science and Technology, Shenzhen 518055, China}
\author{Song Liu}
\email{lius3@sustech.edu.cn}
\affiliation{Shenzhen Institute for Quantum Science and Engineering, Southern University of Science and Technology, Shenzhen 518055, China}
\affiliation{International Quantum Academy, Shenzhen 518048, China}
\affiliation{Guangdong Provincial Key Laboratory of Quantum Science and Engineering, Southern University of Science and Technology, Shenzhen 518055, China}
\author{Youpeng Zhong}
\email{zhongyp@sustech.edu.cn}
\affiliation{Shenzhen Institute for Quantum Science and Engineering, Southern University of Science and Technology, Shenzhen 518055, China}
\affiliation{International Quantum Academy, Shenzhen 518048, China}
\affiliation{Guangdong Provincial Key Laboratory of Quantum Science and Engineering, Southern University of Science and Technology, Shenzhen 518055, China}
\author{Luming Duan}
\email{lmduan@tsinghua.edu.cn}
\affiliation{Center for Quantum Information, Institute for Interdisciplinary Information Sciences, Tsinghua University, Beijing 100084, PR China}
\author{Dapeng Yu}
\email{yudp@sustech.edu.cn}
\affiliation{Shenzhen Institute for Quantum Science and Engineering, Southern University of Science and Technology, Shenzhen 518055, China}
\affiliation{International Quantum Academy, Shenzhen 518048, China}
\affiliation{Guangdong Provincial Key Laboratory of Quantum Science and Engineering, Southern University of Science and Technology, Shenzhen 518055, China}
\affiliation{Department of Physics, Southern University of Science and Technology, Shenzhen 518055, China}

\maketitle

\setcounter{equation}{0}
\setcounter{figure}{0}
\setcounter{table}{0}
\setcounter{page}{1}

\renewcommand{\theequation}{S\arabic{equation}}
\renewcommand{\thefigure}{S\arabic{figure}}
\renewcommand{\thetable}{S\arabic{table}}

\section{Experimental setup}\label{sec:setup}
\begin{figure}[H]
  \centering
  \includegraphics[width=0.8\textwidth]{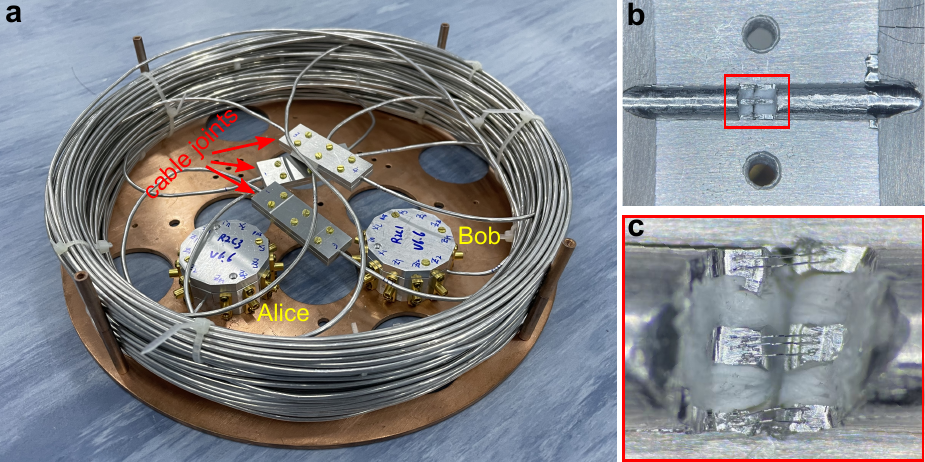}
  \caption{\label{setup}
  {\bf a,} Photograph of the superconducting quantum network assembly.
  {\bf b--c,} Zoomed in micrographs of a cable joint, where two cables are placed end-to-end and clamped on an Al metal jig first, then connected together with bonding wires.
  }
\end{figure}

Figure~\ref{setup}{\bf a} shows a photograph of the superconducting quantum network assembly. The long Al cable is coiled up and mounted on a copper plate with the quantum processors Alice and Bob.
A gradiometer design is followed to reduce its sensitivity to the environmental magnetic field, where half of the cable is coiled clockwise and the other half is coiled counter-clockwise.
Three cable joints (Fig.~\ref{setup}{\bf b--c}) are used to connect four sections of cables each of 16 meters long, thus forming a cable of 64 meters in total.
The cable has an outer conductor diameter of 2.1 mm, an inner conductor diameter of 0.54 mm, and a dielectric insulating layer made of low-density polytetrafluoroethylene (PTFE).
In this setup, Alice and Bob are only separated by about 0.15~m physically, as limited by space inside the fridge; a large physical separation can be achieved with inter-fridge cryogenic link~\cite{Magnard2020}, which is essential for quantum nonlocality tests.

\begin{figure}[H]
  \centering
  \includegraphics[height=0.95\textheight]{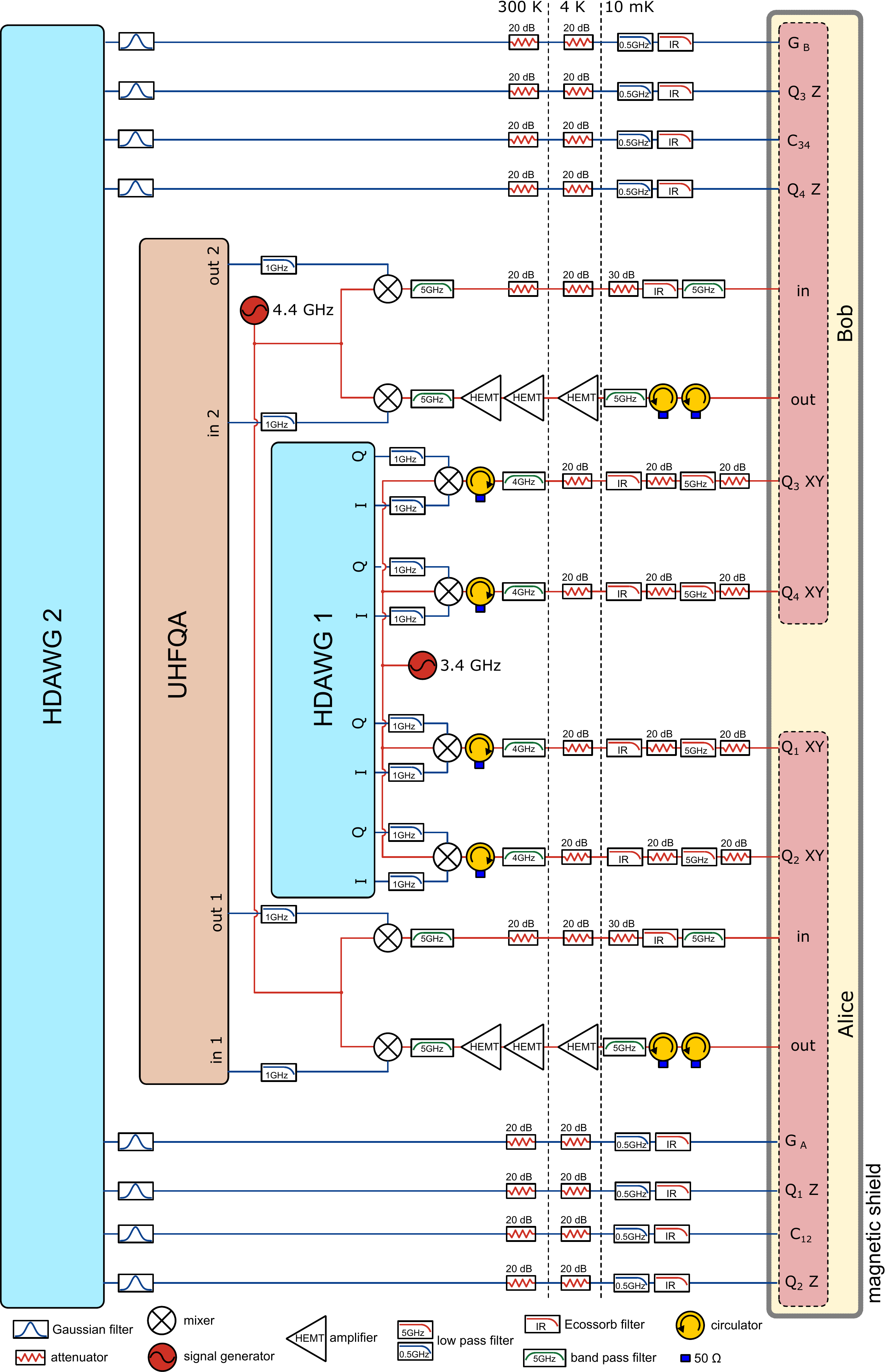}
  \caption{\label{wiring}
  {\bf The cryogenic wiring and room-temperature electronics setup.}}
\end{figure}

A schematic of the cryogenic wiring setup and the room-temperature microwave electronics for qubit control and readout is shown in Fig.~\ref{wiring}.
Commercial electronics manufactured by Zurich Instruments are used in this experiment, where two 8-channel HDAWG modules are used for qubit XY and Z control respectively; an UHFQA module is used for qubit dispersive readout.

An UHFQA has dual intermediate frequency (IF) input/output channels for quadrature modulation/demodulation of one readout chain in a typical setup.
To read out both Alice and Bob with only one UHFQA module, we use a single IF channel and three-port mixer for frequency modulation/demodulation for each readout signal chain.
With an IF sideband of $\sim600$~MHz, the unwanted image sideband is $\sim1.2$~GHz away from the input signal and is rejected by a bandpass filter of 600~MHz bandwidth centered at 5~GHz here.
The output readout signal is down-converted by a three-port mixer and then digitized by the UHFQA module, then demodulated in real time by an FPGA chip in the UHFQA module with sine and cosine integration weights for I and Q quadratures respectively.
A proprietary digital interface (DIO link) is used to feed forward Alice's single-shot qubit measurement result from the UHFQA to HDAWG 1 to trigger the qubit state rotations, with a latency of $\sim 600$~ns (from the end of dispersive readout tone to the start of the rotation pulses, including the time for digital demodulation and qubit state discrimination).

\section{Active cooling of the long cable}\label{sec:thermal}
\begin{figure}[H]
  \centering
  \includegraphics[width=0.8\textwidth]{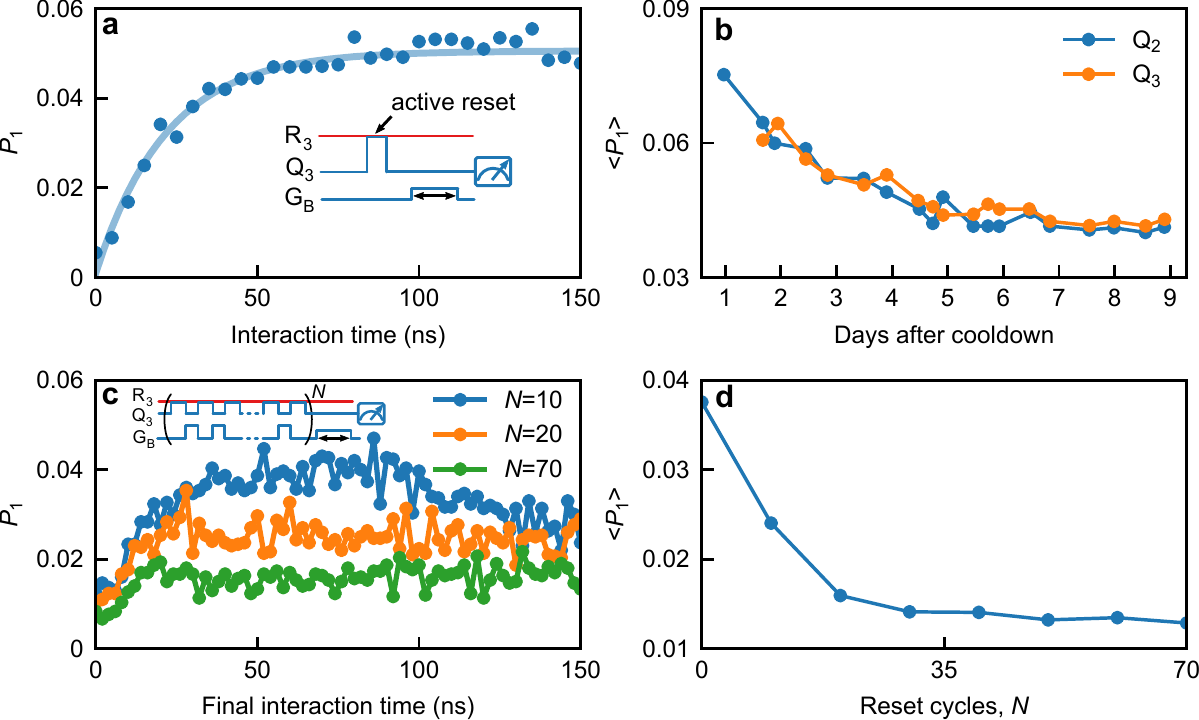}
  \caption{\label{thermal}
  {\bf Active cooling of the long cable.}
  {\bf a,} $Q_3$ interacts with its readout resonator $R_3$ first to cool itself down, then interact with the cable with $G_B$ biased to maximum coupling. The qubit photon population exponentially settles to $\langle P_1\rangle =5\%$ at equilibrium.
  {\bf b,} $\langle P_1\rangle$ as measured in {\bf a} with $Q_2$ and $Q_3$ at different times after the fridge is cooled down, which gradually settles to 4\% after a week.
  {\bf c,} Active cooling of the cable with $Q_3$ by repeatedly swapping the thermal photons out from the cable to the qubit and resetting through the readout resonator. The dynamics of the final interaction is shown, after $N=10$, 20 and 70 cycles of reset through $R_3$.
  {\bf d,} $\langle P_1\rangle$ versus different cycles of active reset $N$. To speed up the active cooling, we use both $Q_2$ and $Q_3$ to swap out the thermal photons in the cable.
}
\end{figure}

To ensure good thermal contact between the long Al cable and the dilution fridge, copper braided wires are used to twine around the cable and then mounted to the mixing chamber stage, the coldest part of the fridge with a base temperature $<10$~mK.
Still, we find that the thermal photon population in the cable is as high as 7.5\% a few hours after cooling down, and gradually settles to 4\% after one week, corresponding to an equilibrium temperature of 60~mK, significantly higher than the fridge base temperature, see Fig.~\ref{thermal}.
Although the source of the thermal noise remains unclear, we find an active reset technique using the qubit readout resonator~\cite{Zhou2021} can reduce the thermal photon population in the qubit to $<0.5$\%; by repeatedly swapping the thermal photons out from the cable to the qubit and actively resetting through the readout resonator, we finally reduce the thermal photon population in the cable from 4\% to 1.5\%, equivalent to cooling from 60~mK to 45~mK, as shown in Fig.~\ref{thermal}.
This active cooling method may also be useful with higher channel temperatures as proposed in Refs.~\onlinecite{Xiang2017,Vermersch2017}.

\section{Device parameters and performance}
The design/measured parameters and typical performance of each qubit and coupler are summarized in Table \ref{parameters}. More details can be found in the subsequent subsections.
\begin{table}[H]
\begin{center}
\begin{tabular}{|c |c c c c c c c c|}
  \hline
  \hline
  parameter & $Q_1$ & $Q_2$ & $Q_3$ & $Q_4$ & $C_{12}$ & $C_{34}$ & $G_A$ & $G_B$  \\
  \hline
  min freq., $\omega_{q}^{\rm{min}}/2\pi$ (GHz)     & 3.778 & 3.926 & 3.935 & 3.643& $\sim 3.2$ &$\sim 3.2$&   &  \\
  max freq., $\omega_{q}^{\rm{max}}/2\pi$ (GHz)     & 4.716 & 4.771 & 4.898 & 4.636& $\sim 6.8$ &$\sim 6.8$&   &  \\
  idling freq., $\omega_{q}/2\pi$ (GHz)             & 3.7894 & 3.9331 & 3.9356&3.6430 & $\sim 6.8$ & $\sim 6.8$&  &  \\
  nonlinearity, $\eta_q/2\pi$ (MHz)                 & $-207$ & $-191$ & $-194$&$-212$& $\sim -400$ &$\sim -400$&   &  \\
  lifetime, $T_1$ ($\mu$s)                           & 37.1 & 34.7 & 33.6 &20.1&     &   & &  \\
  dephasing time, $T_\phi$ ($\mu$s)                  & 22.8 & 26.3 & 62.6 &33.6&     &   &  &\\
  readout resonator freq., $\omega_{rr}/2\pi$ (GHz)  & 4.9153 & 4.9562 & 4.9559&4.9153 &     & &   &  \\
  readout time ($\mu$s)                              & 0.7 & 0.7 & 0.7 & 0.7&    &   & & \\
  $|0\rangle$ readout fidelity, $F_{0}$          & 99.6\% & 99.2\% & 99.3\% &99.6\%&   &  &   &  \\
  $|1\rangle$ readout fidelity, $F_{1}$          & 98.0\% & 97.7\% & 97.3\% &97.7\%&   &  &   &  \\
  max coupling, $\kappa_n^{\rm{max}}$ (ns$^{-1}$)                  &  &  &  & &   & & $1/22$  & $1/18$ \\
  \hline
  \hline
\end{tabular}
\end{center}
\caption{\label{parameters} {\bf Device parameters and typical performance.}}
\end{table}

\subsection{Fast and high fidelity single-shot qubit readout}\label{sec:readout}
\begin{figure}[H]
  \centering
  \includegraphics[width=0.8\textwidth]{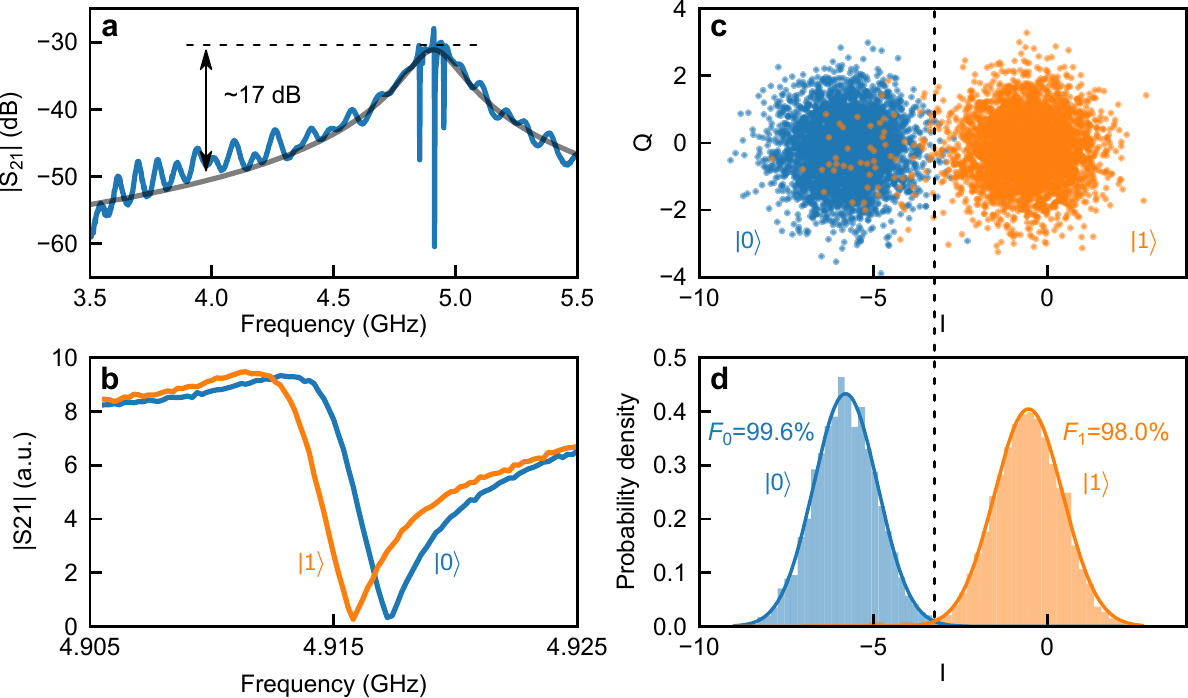}
  \caption{\label{dispersive_readout}
  {\bf Single-shot qubit dispersive readout with Purcell filter.}
  {\bf a,} Transmission spectrum of the Purcell filter. The readout resonators are in the passband, whereas the qubit frequencies are 1 GHz detuned, with $\sim$17~dB rejection ratio.
  {\bf b,} Dispersive shift of the readout resonator depending on the qubit state of $Q_1$.
  {\bf c,} Single-shot dispersive readout in the quadrature (IQ) space. Each data point represents the demodulation result of the 0.7 $\mu$s long probe microwave signal, with state dependent phase shift imparted by the qubit.
  {\bf d,} Histogram of the result in {\bf c} projected to the $I$ quadrature. The dashed line is the state discrimination threshold, yielding a $|0\rangle$ state fidelity of $F_0=99.6\%$ and a $|1\rangle$ state fidelity of $F_1=98.0\%$. Solid lines are numerical fits to Gaussian distribution, where the $|0\rangle$ state separation error is 0.28\%, suggesting a 0.12\% residual thermal population in the qubit.
  }
\end{figure}

A fast and high fidelity single-shot qubit readout is essential for teleportation with feed-forward.
However, these two figures-of-merit have different, and sometimes competing requirements. Simultaneous satisfaction of both requires careful design of system parameters and the assistance of additional elements~\cite{Walter2017}.
For transmon qubits, the qubit state is measured by probing the state dependent frequency shift (dispersive shift) of a coupled but largely detuned resonator, but spontaneous decay to the readout feed line through the resonator leads to extra qubit damping via the Purcell effect~\cite{Houck2008}.
One way to suppress this energy relaxation through the readout resonator is to employ a bandpass filter between the resonator and the environment which impedes microwave propagation at the qubit frequency, suppressing the Purcell rate without compromising qubit readout speed.
Adopting the Purcell filter designs from Refs.~\onlinecite{Jeffrey2014,Satzinger2018,Bienfait2019,Zhong2021}, we use a short-circuited, half-wavelength ($\lambda/2$) coplanar waveguide (CPW) resonator as a single-pole bandpass filter, where the input and output lines intersect with the CPW line near the shorted ends.
This Purcell filter has a center frequency of 4.91~GHz, a weak coupling to the input port (coupling $Q_c\sim2000$) and a strong coupling to the output port (coupling $Q_c\sim25$), see its transmission spectrum in Fig.~\ref{dispersive_readout}{\bf a}.
With this Purcell filter, we are able to perform high-fidelity, single-shot qubit dispersive readout even absent a parametric amplifier~\cite{Vijay2011}, see Fig.~\ref{dispersive_readout}{\bf b-d}.

\begin{figure}[H]
  \centering
  \includegraphics[width=0.9\textwidth]{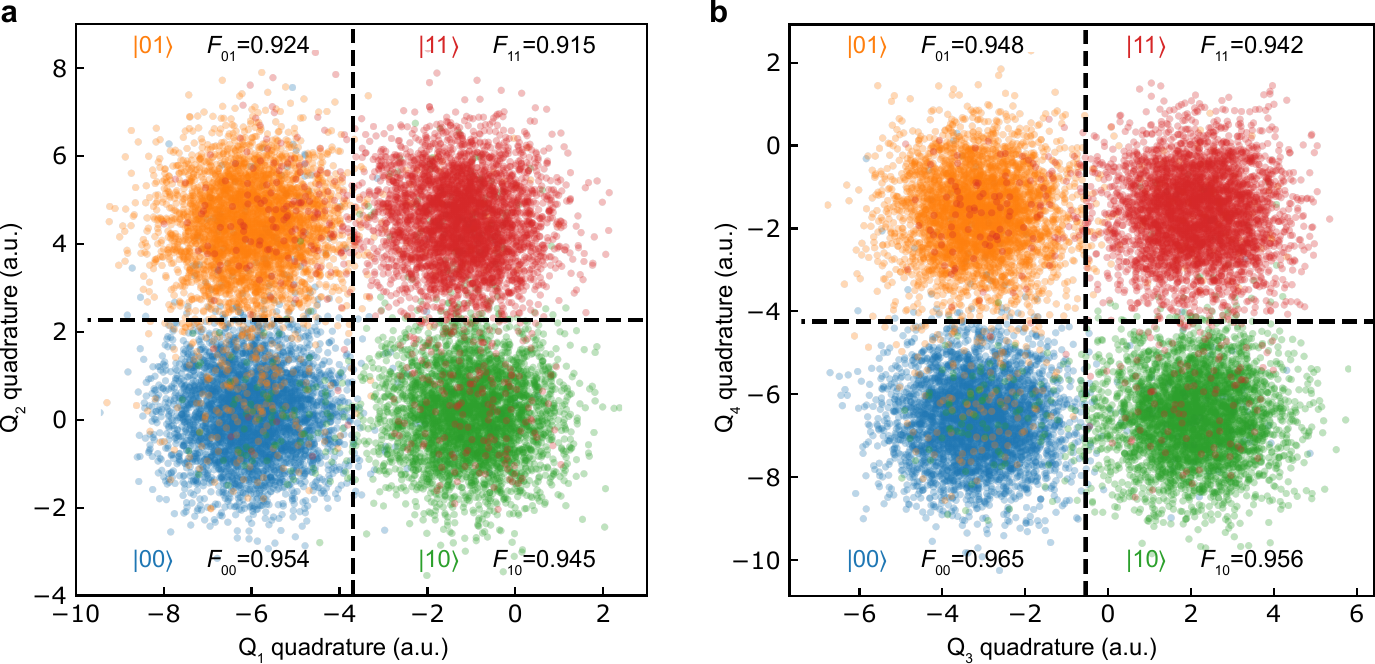}
  \caption{\label{Q1Q2Q3Q4_readout} {\bf Joint readout of $Q_1$-$Q_2$ and $Q_3$-$Q_4$.} }
\end{figure}

Using frequency multiplexing, we simultaneously probe the readout resonators of $Q_1$ and $Q_2$ ($Q_3$ and $Q_4$). Each qubit imparts a state dependent phase shift to the corresponding frequency component of the measurement pulse. By demodulating the measurement pulse at the readout frequencies of $Q_1$ and $Q_2$ ($Q_3$ and $Q_4$) and projecting to proper quadratures respectively, we obtain the joint readout result of both qubits, see Fig.~\ref{Q1Q2Q3Q4_readout}. The dashed lines are state discrimination thresholds, with a $|00\rangle$ state fidelity of $F_{00}=95.4\%$, a $|01\rangle$ state fidelity of $F_{01}=92.4\%$, a $|10\rangle$ state fidelity of $F_{10}=94.5\%$ and a $|11\rangle$ state fidelity of $F_{11}=91.5\%$ for $Q_1$-$Q_2$ joint readout, and  $F_{00}=96.5\%$, $F_{01}=94.8\%$, $F_{10}=95.6\%$ and $F_{11}=94.2\%$ for $Q_3$-$Q_4$ joint readout, respectively.

\subsection{Qubit-qubit coupling}\label{sec:Tcoupler}
\begin{figure}[H]
  \centering
  \includegraphics[width=0.9\textwidth]{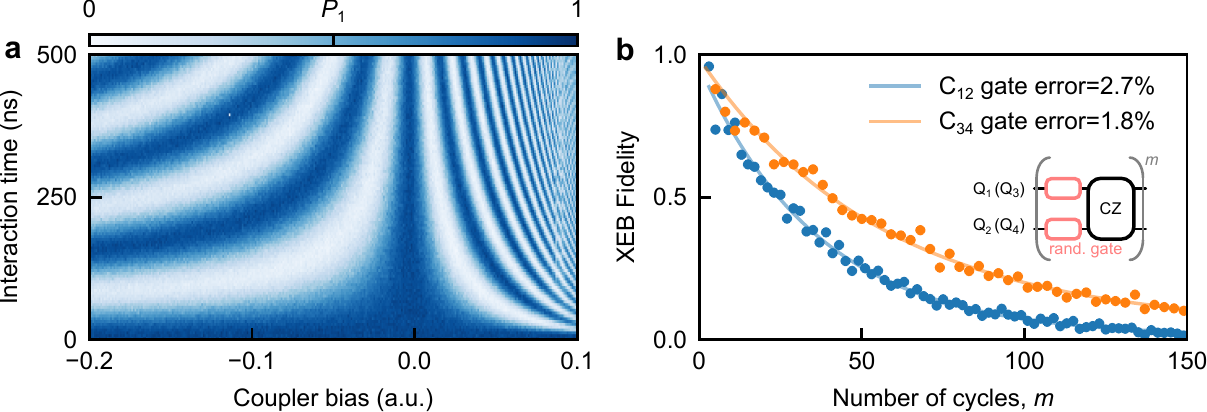}
  \caption{\label{Tcoupler} {\bf Characterization of the T-shaped transmon couplers $C_{12}$ and $C_{34}$.}
  {\bf a,} Vacuum Rabi oscillation between $Q_1$ and $Q_2$ at different coupler bias.
  {\bf b,} Cross-entropy benchmarking (XEB) of the local CZ gate between $Q_1$ and $Q_2$ ($Q_3$ and $Q_4$). The $Q_1$-$Q_2$ CZ gate fidelity is 97.3\%, $Q_3$-$Q_4$ CZ gate fidelity is 98.2\%.}
\end{figure}
In Alice (Bob), $Q_1$ and $Q_2$ ($Q_3$ and $Q_4$) are coupled to each other with a transmon style, T-shaped coupler $C_{12}$ ($C_{34}$)~\cite{Yan2018}. The coupler can mediate a coupling strength from 3~MHz to $-25$~MHz continuously, see Fig.~\ref{Tcoupler}{\bf a}.
The CZ gate is implemented in a non-adiabatic manner as Ref.~\onlinecite{Sung2021}. We use a square pulse to bring the states $|11\rangle$ and $|02\rangle$ into resonance, where the $|11\rangle$ state acquires a phase $\pi$ after 90~ns of swapping, here $|2\rangle$ is the second excited state of the qubit.
we use the cross-entropy benchmarking (XEB) technique~\cite{Arute2019} to benchmark the CZ gates, as shown in Fig.~\ref{Tcoupler}{\bf b}, where the $Q_1$-$Q_2$ CZ gate fidelity is 97.3\%, the $Q_3$-$Q_4$ CZ gate fidelity is 98.2\%.



\subsection{Qubit-cable coupling}\label{sec:mmode}
\subsubsection{Jaynes-Cummings model}\label{sec:JC}
At weak multimode coupling regime, the long cable can be treated as a series of multimode resonators, and the $Q_2$-cable-$Q_3$ system can be modeled with the following rotating-frame Hamiltonian:
\begin{eqnarray}\label{H}
  H/\hbar &=& \sum_{i=2,3} \Delta\omega_{q,i} \sigma_{i}^\dag \sigma_{i} + \sum_{m} \Delta \omega_{m} a_m^\dag a_m \\
  &&+\sum_{m} g_{2,m} \left (\sigma_{2} a_m^\dag + {\sigma_{2}}^\dag a_m \right ) +\sum_{m} (-1)^m g_{3,m} \left (\sigma_{3} a_m^\dag + {\sigma_{3}}^\dag a_m \right)\nonumber,
\end{eqnarray}
where $\sigma_i$ and $a_m$ are the annihilation operators for the two qubits and the $m$-th standing wave mode respectively, $\Delta\omega_{q,i}$ and $\Delta\omega_m$ are the qubit and standing wave mode frequency detuning with respect to the rotating frame frequency.
Note the sign of $g_{3,m}$ alternates with the mode number $m$ due to the parity change of the standing wave mode~\cite{Pellizzari1997,Vogell2017}.

The Jaynes-Cummings coupling strength $g_{i,m}$ between $Q_i$ and the $m$-th mode can be determined from a lumped element circuit model.
The $\ell_{cb}=64$~m long Al cable has a specific capacitance $\mathscr{C}_{cb} = 86.5$~pF/m provided by the cable manufacturer.
From the speed of light $v_c\approx 2.4\times 10^8$~m/s in the cable we estimate a specific inductance $\mathscr{L}_{cb} \approx 200$~nH/m.
The $m$-th standing wave mode in the cable can be modeled as a lumped element series $LC$ resonator~\cite{Pozar}, with parameters given by
\begin{eqnarray}
  L_m &\approx& \frac{1}{2} \mathscr{L}_{cb} \ell_{cb} = {\rm6400\: nH}, \label{Lm}\\
  \omega_m &\approx& m \omega_{\rm FSR},\\
  C_m &=& \frac{1}{\omega_m^2L_m}.
\end{eqnarray}

Following Refs.~\onlinecite{Chen2014,Zhong2019,Zhong2021}, the coupling between $Q_i$ and the cable via the coupler $G_n$ ($n=A$, $B$) is modeled by a tunable mutual inductance given by~\cite{Chen2014,Geller2015}
\begin{equation}\label{M}
  M_n = \frac{L_{g}^2}{2L_{g}+L_w+L_{T,n}/\cos\delta_n},
\end{equation}
where $\delta_n$ is the phase across the Josephson junction of $G_n$, $L_{T,n}$ is the coupler junction inductance at $\delta_n=0$, $L_w \approx 0.06$ nH represents the stray wiring inductance of the CPW line connecting the junction with the two linear inductors $L_{g}=0.2$~nH, which cannot be ignored when $L_{T,n}$ becomes very close to $2L_{g}$~\cite{Zhong2019}.

In the harmonic limit and assuming weak coupling, the Jaynes-Cummings coupling strength $g_{i,m}$ is~\cite{Chen2014,Geller2015}
\begin{equation}\label{coupling_m}
  g_{i,m} = -\frac{M_n}{2} \, \sqrt{\frac{\omega_m \omega_{q,i}}{(L_{g}+L_{J,i})(L_{g}+L_m)}} \approx -\frac{M_n}{2} \, \sqrt{\frac{\omega_m \omega_{q,i}}{(L_{g}+L_{J,i})L_m}},
\end{equation}
where $L_{J,i}$ is the qubit $Q_i$'s junction inductance and $\omega_{q,i}/2\pi$ is $Q_i$'s operating frequency, and $L_g$ can be ignored compared to $L_m$.

By varying the tunable coupler flux bias $\Phi_n$, the junction phase $\delta_n$ is changed as~\cite{Geller2015}
\begin{equation}\label{delta_Phi}
2\pi \frac{\Phi_n}{\Phi_0} = \delta_n + \frac{2L_g+L_w}{L_{T,n}} \sin \delta_n,
\end{equation}
where $\Phi_0=2.067\times 10^{-15}$~Wb is a flux quantum.
At $\Phi_n=\pm \Phi_0/2$, $\delta_n = \pm \pi$, Eq.~\ref{M} reaches maximum, so as the coupling strength.
To turn off the coupling, we should set $\delta_n=\pi/2$, where the flux bias $\Phi_{\rm{off}}$ satisfies
\begin{equation}\label{delta_Phi}
2\pi \frac{\Phi_{\rm{off}}}{\Phi_0} = \frac{\pi}{2} + \frac{2L_g+L_w}{L_{T,n}}.
\end{equation}

\subsubsection{Input-output theory}\label{sec:input_output}
When the qubit-cable coupling becomes strong enough, the microwave photons emitted from the qubit have temporal widths comparable to the cable length, in this regime, it is more appropriate to treat this ``ping-pong'' dynamics in the time domain.
In the single-excitation manifold where the qubit nonlinearity can be ignored, we can simply treat the qubits as quantum harmonic oscillators, and the flying photon transfer process can be modeled with input-output theory~\cite{Gardiner1985,Cirac1997,Bienfait2019,Zhong2019}:
\begin{eqnarray}
  \frac{\rm{d} \sigma_2}{\rm{d}t} &=& -i\Delta\omega_{q,2} \sigma_2 -\frac{\kappa_A(t)}{2} \sigma_2 + \sqrt{\kappa_A(t)} a_{\rm{in},2}(t), \label{2io1}\\
  \frac{\rm{d} \sigma_3}{\rm{d}t} &=& -i\Delta\omega_{q,3} \sigma_3 -\frac{\kappa_B(t)}{2} \sigma_3 + \sqrt{\kappa_B(t)} a_{\rm{in},3}(t), \label{2io2}
\end{eqnarray}
here $a_{{\rm{in}},i}$ ($i=2$, 3) is the input field operator given by
\begin{eqnarray}
  a_{\rm{in},2}(t) &=& \sqrt{\eta} a_{\rm{out},3}(t-\tau_{st}),\label{2io5} \\
  a_{\rm{in},3}(t) &=& \sqrt{\eta} a_{\rm{out},2}(t-\tau_{st}),\label{2io6}
\end{eqnarray}
and the output field operator $a_{{\rm{out}},i}$ is determined by
\begin{eqnarray}
a_{\rm{out},2}(t) &=& \sqrt{\kappa_A(t)} \sigma_{2}(t) - a_{\rm{in},2}(t),\label{2io3} \\
a_{\rm{out},3}(t) &=& \sqrt{\kappa_B(t)} \sigma_{3}(t) - a_{\rm{in},3}(t).\label{2io4}
\end{eqnarray}
The photon emission rates $\kappa_n$ can be calculated from the on-resonant Jaynes-Cummings coupling strength $g_{i,m}$ using Fermi's golden rule:
 \begin{equation}\label{Fermi}
   \kappa_n = \frac{2\pi}{\hbar} (\hbar g_{i,m})^2\frac{1}{\hbar \omega_{\rm{FSR}}} = \frac{2\pi g_{i,m}^2}{\omega_{\rm{FSR}}}.
 \end{equation}
Inserting Eq.~\ref{coupling_m} into Eq.~\ref{Fermi} with $\omega_m = \omega_{q,i}$, and remember
\begin{equation}\label{fsr}
  \ell_{cb} = \tau_{st}  v_c = \frac{\pi}{\omega_{\rm{FSR}}} \frac{1}{\sqrt{\mathscr{L}_{cb} \mathscr{C}_{cb}}},
\end{equation}
we have
 \begin{equation}\label{Fermi}
   \kappa_n = \frac{(M_n \omega_{q,i})^2}{(L_{J,i}+L_g) Z_0} \propto M_n^2,
 \end{equation}
where $Z_0 = \sqrt{\mathscr{L}_{cb}/\mathscr{C}_{cb}}$ is the characteristic impedance of the cable.

\begin{figure}[h]
  \centering
  \includegraphics[width=3in]{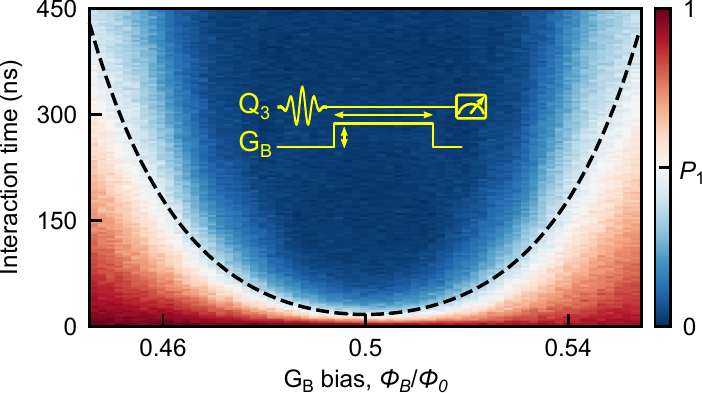}\\
  \caption{\bf Characterizing the emission rate of $G_B$ at different bias.}\label{fig3aQ3}
\end{figure}

In Fig.~3{\bf a} in the main text we have characterized the emission rate of $G_A$, here we show the data for $G_B$ in Fig.~\ref{fig3aQ3}, with a maximum coupling rate of $\kappa_B^{\rm max}=1/18$~ns$^{-1}$.

\begin{figure}[H]
\begin{center}
	\includegraphics[width=0.85\textwidth]{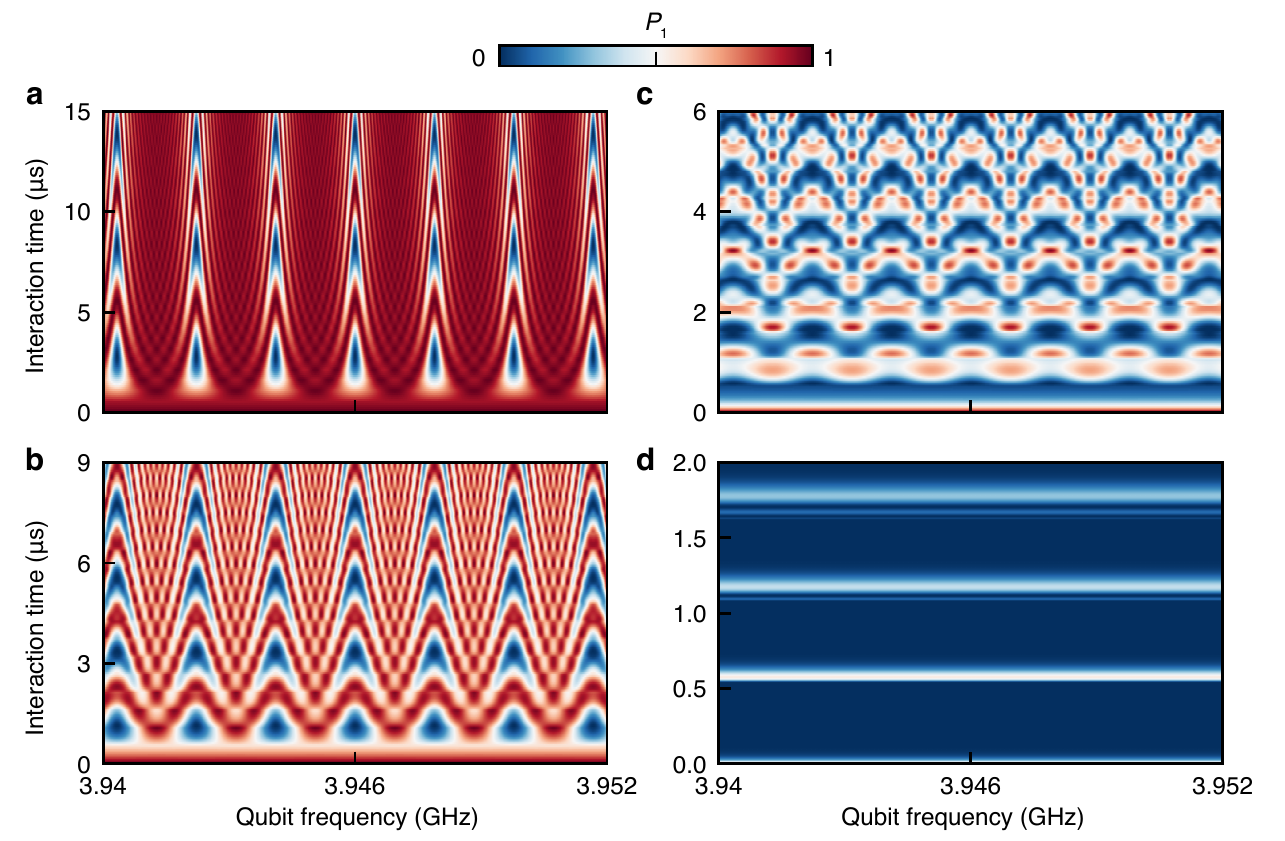}
	\caption{
    \label{fig2simu}
    {\bf Simulation of the transition from stationary to flying microwave photons.}
    }
\end{center}
\end{figure}

In Fig.~\ref{fig2simu}, we simulate the transition from stationary to flying microwave photons in Fig.~2 in the main text.
Fig.~\ref{fig2simu}{\bf a}-{\bf c} is calculated with the Jaynes-Cummings multimode Hamiltonian in Eq.~\ref{H}, without taking decoherence into account to save numerical computation time.
Fig.~\ref{fig2simu}{\bf d} is calculated using the input-output theory model Eq.~\ref{2io1}. The tiny emerging stripes in Fig.~2{\bf d} is also captured in the simulation in Fig.~\ref{fig2simu}{\bf d}, which arise from the interference between the wavepacket that is immediately reflected by the qubit and the wavepacket that is captured and re-emitted by the qubit.

 \subsubsection{Shaping photons}\label{sec:shape}
 Naturally emitted photons have exponentially decaying envelopes, for which a receiver with fixed coupling can have at most 54\% absorption efficiency~\cite{Stobinska2009,Wang2011}.
 As mentioned in the main text, by dynamically tuning the sender coupling to symmetrize the emitted photon envelope, and tuning the receiver coupling reversely, 100\% transfer efficiency can be achieved because of time-reversal-symmetry~\cite{Cirac1997,Korotkov2011}. There exist different solutions of $\kappa_A (t)$ that symmetrize $a_{\rm{out},2}(t)$ in Eq.~\ref{2io3}, here we follow Refs.~\onlinecite{Kurpiers2018,Bienfait2019} and dynamically tune the coupling as:
\begin{eqnarray}
   \kappa_A (t) &=& \kappa_c \frac{e^{\kappa_c t}}{1 + e^{\kappa_c t}}, \label{kappaA}\\
   \kappa_B (t) &=&  \kappa_A (\tau_{st} - t), \label{kappaB}
\end{eqnarray}
where $\kappa_c$ is the characteristic emission rate during this process.
In the absence of any incoming field, the emitted photon has a hyperbolic secant envelope $a_{\rm{out},2}(t) \propto \rm{sech}(\kappa_c t/2)$.

To emit a fraction of a photon, for example half a photon for creating remote entanglement, we can tune the emission rate as
\begin{eqnarray}\label{fraction_photon}
   \kappa_A (t) = \kappa_c \frac{\alpha}{1+ (1-\alpha) e^{\kappa_c t} } \frac{e^{\kappa_c t}}{1 + e^{\kappa_c t}},
\end{eqnarray}
where $\alpha$ is the fraction of a photon been emitted.
Note the receiver coupling should still be tuned as the reverse of Eq.~\ref{kappaA} instead of Eq.~\ref{fraction_photon}.

\begin{figure}[H]
  \centering
  \includegraphics[width=0.8\textwidth]{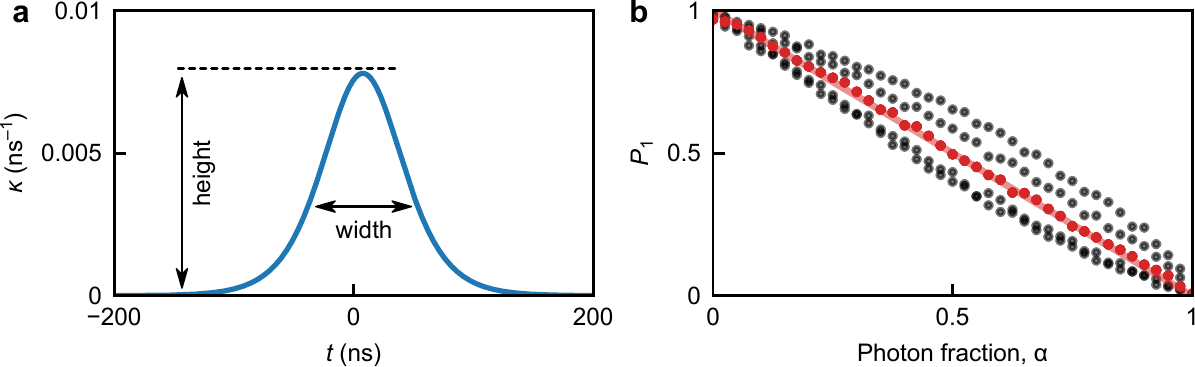}
  \caption{\label{shape_photon} {\bf Photon shaping calibration.}
  {\bf a}, Dynamic tuning of $\kappa$ with $\alpha=0.5$ and $\kappa_c=1/22$~ns, showing two key features, the coupling height determined by $\kappa_c$, a time-independent value, and the temporal width, which should match $\kappa_c$.
  {\bf b}, Photon emission performed with different $\alpha$. Ideally, the population $P_1$ remained in the sender qubit should be $1-\alpha$. Deviation from this linear curve suggests mismatch between the pulse height and width in {\bf a}, where a convex curve shape suggests under coupling (actual coupling height smaller than target), and a concave curve shape suggests over coupling (actual coupling height larger than target). The data in red represents a good calibration result.}
\end{figure}

Shaping the emitted photon envelope precisely is crucial for efficient transfer of flying photons, but also very challenging experimentally.
Controlled emission of microwave photons has been first explored on single chips using different methods~\cite{Yin2013,Srinivasan2014,Pechal2014}; subsequent two-node experiments~\cite{Kurpiers2018,Axline2018,Campagne2018,Magnard2020} use a circulator to interrupt the communication channel, opening a window to monitor the photon envelope for control pulse calibration, but at the cost of extra channel loss.
Here, without a circulator interrupting the channel for monitoring, we need to find a sensitive indicator for calibration.
The photon emission experiment in Fig.3{\bf a} in the main text provides a good calibration for $\kappa_A$ at different gmon coupler flux bias, but only in a static manner.
We find this static calibration result is not precise enough when shaping the photon envelopes dynamically.
It turns out that Eq.~\ref{fraction_photon} provides a method to calibrate $\kappa_A$ precisely in a dynamic manner.
A representative curve of Eq.~\ref{fraction_photon} is shown in Fig.~\ref{shape_photon}{\bf a} with $\alpha=0.5$, showing two key features, the pulse height determined by $\kappa_c$, a time-independent value, and the temporal width, determined by the time-dependent part $\frac{\alpha}{1+ (1-\alpha) e^{\kappa_c t} } \frac{e^{\kappa_c t}}{1 + e^{\kappa_c t}}$.
As we vary $\alpha$ and perform photon emission, the population $P_1$ remained in the sender qubit should be $1-\alpha$ ideally, regardless of the photon capture process.
Deviation from this linear curve suggests that the pulse height and width are not matched, see Fig.~\ref{shape_photon}{\bf b}. We then adjust our calibration until $P_1$ is well aligned with expectation, as shown by the red data in Fig.~\ref{shape_photon}{\bf b}.

\subsubsection{Frequency compensation}\label{sec:freq_comp}
  An ideal coupler should only change the coupling strength and not affect the qubit.
  However, varying the gmon coupler junction inductance $L_{T,n}/\cos\delta_n$ also affects the qubit inductance via the mutual inductance:
\begin{equation}\label{Lq}
  L_{q,i} = L_{J,i} + L_g - M_n,
\end{equation}
This inductance variation induces a qubit frequency shift~\cite{Zhong2019}
\begin{equation}\label{delta_fq}
  \delta \omega_{q,i} = - g_{i,m} \sqrt{ \frac{L_g + L_m}{L_g+L_{J,i}}}\approx  - g_{i,m} \sqrt{ \frac{L_m}{L_g+L_{J,i}}}.
\end{equation}
We can also express the frequency shift in terms of $\kappa_n$ using Eq.~\ref{coupling_m}:
\begin{equation}\label{delta_fq2}
  \delta \omega_{q,i} = - \sqrt{\frac{g_{i,m}^2 L_m}{L_g+L_{J,i}}} = - \sqrt{\frac{\kappa_n \omega_{\rm{FSR}}}{2\pi}  \frac{L_m}{L_g+L_{J,i}}} ,
\end{equation}
inserting Eq.~\ref{Lm} and Eq.~\ref{fsr} into Eq.~\ref{delta_fq2}, we have
\begin{equation}\label{delta_fq3}
  \delta \omega_{q,i} = - \frac{1}{2} \sqrt{\frac{\kappa_n \sqrt{\mathscr{L}_{cb}/\mathscr{C}_{cb}}}{(L_g+L_{J,i})}} = - \frac{1}{2} \sqrt{\frac{\kappa_n Z_0}{(L_g+L_{J,i})}} \propto \sqrt{\kappa_n}.
\end{equation}

\begin{figure}[H]
  \centering
  \includegraphics[width=3.5in]{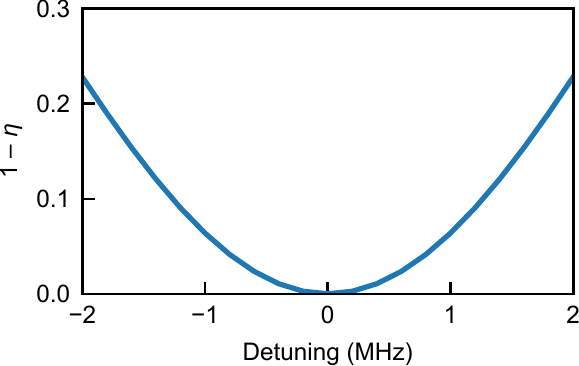}
  \caption{\label{freq_comp} {\bf Flying photon transfer inefficiency versus frequency mismatch.} Numerical simulation suggests that 2~MHz frequency mismatch between $Q_2$ and $Q_3$ could cause 22\% transfer inefficiency.
  }
\end{figure}

Theoretical studies show that the flying photon transfer process is very sensitive to frequency mismatch between the sender and receiver~\cite{Sete2015}.
For example, with the parameters in our experiment, we find that a frequency mismatch of merely 2~MHz between $Q_2$ and $Q_3$ could cause 22\% photon transfer inefficiency, see Fig.~\ref{freq_comp}{\bf a}.
Therefore a precise active frequency compensation is essential for efficient transfer of flying photons.

The negative sign in Eq.~\ref{delta_fq3} suggests that the qubit frequency is decreased as the coupling rate is increased, as clearly shown in Fig.~3{\bf b} in the main text. In this experiment, we use asymmetric Josephson junctions with $\alpha=E_{J1}/E_{J2}=4.7$, where $E_{J1}$ and $E_{J2}$ are the Josephson energies of the two qubit junctions~\cite{Hutchings2017}. With this configuration, the qubits have a maximum frequency sweet spot near 5~GHz and a minimum frequency sweet spot near 3.9~GHz. it is convenient to operate the qubits near their minimum frequency sweet spots, where they have long coherence times and the negative frequency shift can be easily compensated with qubit Z bias.

\subsubsection{Transfer inefficiency analysis}\label{sec:inefficiency}
In the main text, we achieve a single photon transfer efficiency of 90.4\%.
the loss in the cable contributes $1-\exp(-\tau_{st}/T_1^m)=0.5\%$ inefficiency, the rest 9.1\% inefficiency mainly comes from residual thermal photons in the cable, qubit decoherence, weak reflections at the cable joints and control pulse imperfection, including frequency mismatch and pulse shape asymmetry~\cite{Sete2015}. The 1.5\% residual thermal photon population in the cable is estimated to contribute 1.3\% inefficiency, following the method in Refs.~\onlinecite{Xiang2017,Vermersch2017}. The qubit decoherence contributes about 1.2\% inefficiency, by adding a self-decaying term $-\frac{\sigma_i}{2T_{1,i}}$ to Eqs.~\ref{2io1} and \ref{2io2} respectively. The contribution of the weak reflections at the cable joints is estimated using numerical simulation with HFSS, as provided below. Our model suggests that the weak reflections at the cable joints contribute $\sim1$\% inefficiency. The rest 5.6\% inefficiency is believed to arise from control pulse imperfection, including frequency mismatch and pulse shape asymmetry.

\begin{figure}[H]
  \centering
  \includegraphics[width=0.9\textwidth]{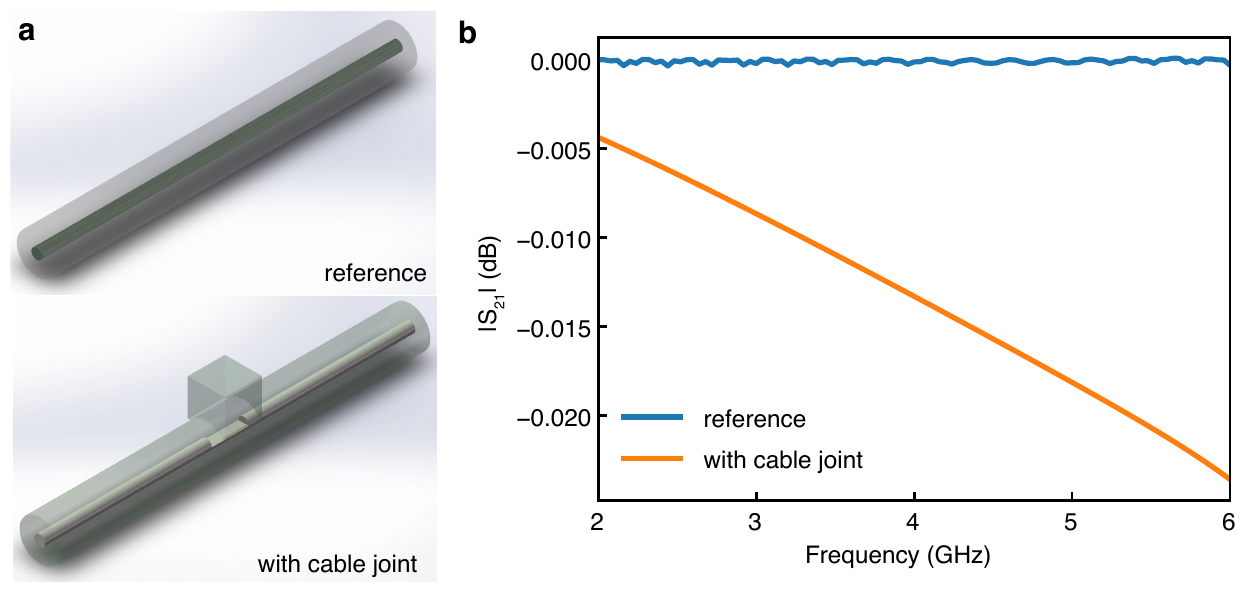}
  \caption{\label{HFSS} {\bf Simulating the cable joint transmission.}
  {\bf a} Schematic of the HFSS model that simulates the transmission of a cable joint. The top picture shows a reference coaxial cable model, the bottom picture shows a model with a simplified cable joint.
  {\bf b} Simulated $S_{21}$ parameter for the two models in {\bf a}. $|S_{21}|=-0.015$~dB for a cable joint at around 4~GHz.
  }
\end{figure}

It is technically challenging to accurately estimate the transmission and reflection parameters of a cable joint.
Here we build a simplified model with HFSS and simulate the scattering parameters of a cable joint, see Fig.~\ref{HFSS}{\bf a}.
The transmission parameter $|S_{21}|=-0.015$~dB for a cable joint at around 4~GHz.
Ignoring the cascaded reflections, three cable joints roughly have a transmission of $-0.045$~dB, equivalent to 1\% photon inefficiency.

\section{Quantum state and process tomography}\label{sec:tomo}
\begin{figure}[H]
  \centering
  \includegraphics[width=0.9\textwidth]{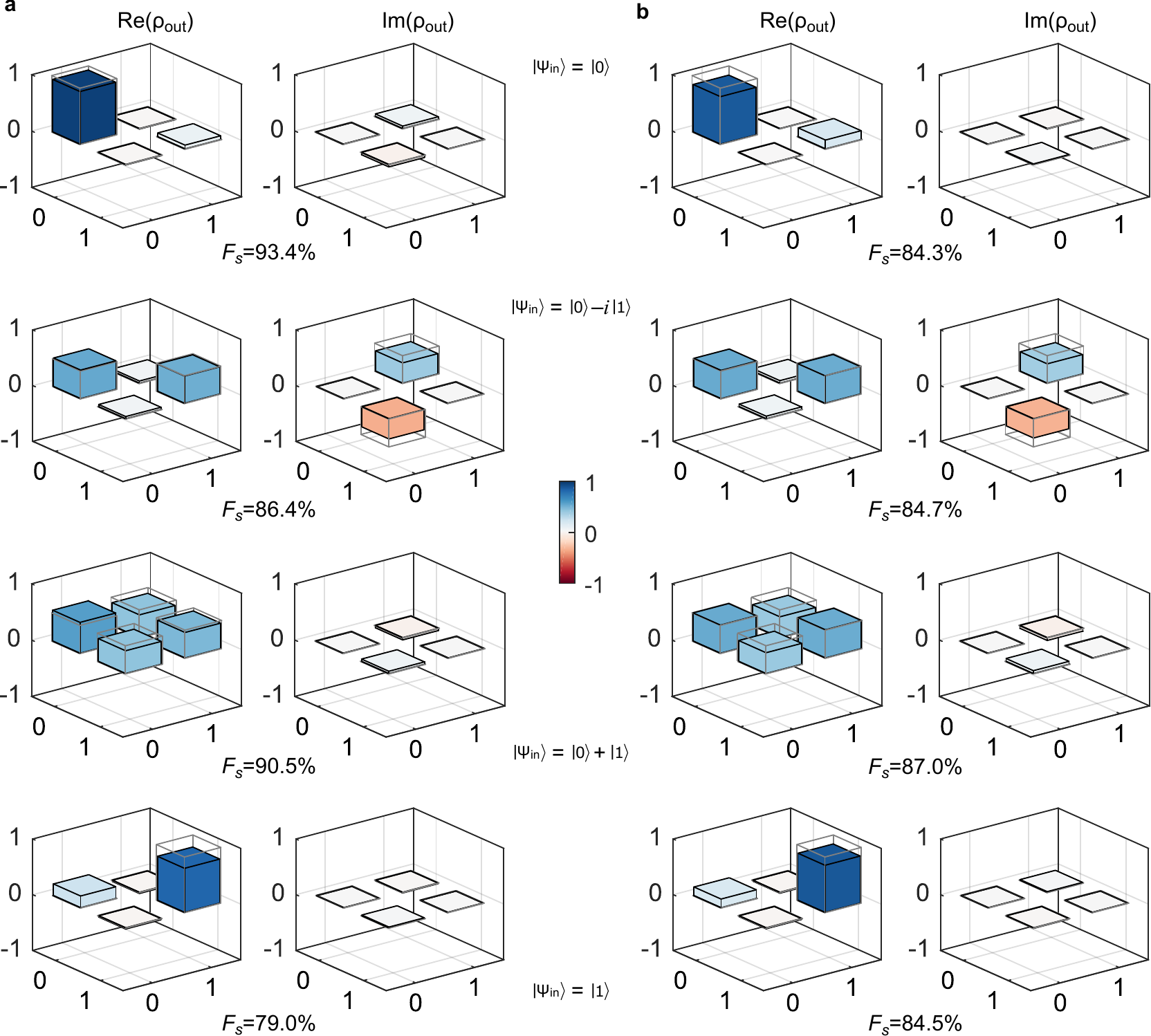}
  \caption{\label{teleport} {\bf Quantum state tomography of the teleported output states.}
  {\bf a,} Post-selected $Q_3$ output state density matrix $\rho_{\rm out}$ conditioned upon a $Q_1$-$Q_2$ joint measurement outcome of $ij=00$, with state fidelities of $F_s=93.4\%$, 86.4\%, 90.5\% and 79.0\% for the four input states $|0\rangle$, $|0\rangle-i|1\rangle)$, $|0\rangle+|1\rangle$ and $|1\rangle$ respectively.
  {\bf b,} $Q_3$ output state density matrix $\rho_{\rm out}$ of the deterministic teleportation with feed-forward, with state fidelities of $F_s=84.3\%$, 84.7\%, 87.0\% and 84.5\% for the four input states respectively.
  The solid bars and grey frames are the measured and ideal values respectively.
  }
\end{figure}

Quantum state tomography~\cite{Steffen2006} is used to characterize the teleported output quantum states. We prepare the input state $|\psi_{\rm{in}}\rangle$ of $Q_1$ in $|0\rangle$, $|0\rangle-i|1\rangle$, $|0\rangle+|1\rangle$ and $|1\rangle$, then perform the quantum state teleportation, then apply tomography pulses $\{I, X/2, Y/2\}$ to $Q_3$ before the readout pulse. The output state density matrix $\rho_{\rm out}$ is reconstructed from the measured probability outcome, see Fig.~\ref{teleport}{\bf a} and {\bf b} for the reconstructed density matrices of different input states, using post-selection and feed-forward respectively.
Quantum process tomography~\cite{Neeley2008} for the teleportation is performed using $\{|0\rangle$, $|0\rangle-i|1\rangle$, $|0\rangle+|1\rangle$, $|1\rangle\}$ as input states and the measured $Q_3$ density matrices in Fig.~\ref{teleport} as output states.
The single-shot readout is repeated $4096$ times to obtain the measured probabilities; the state and process tomography is run repeatedly, in each repeat we reconstruct the density matrix and process matrix. The fidelities and uncertainties of the teleportation correspond to the mean and standard deviation of 40 repeated measurements, corresponding to $1.6\times 10^5$ single-shot experiments, costing about 3.5 hours.

\begin{figure}[H]
  \centering
  \includegraphics[width=0.9\textwidth]{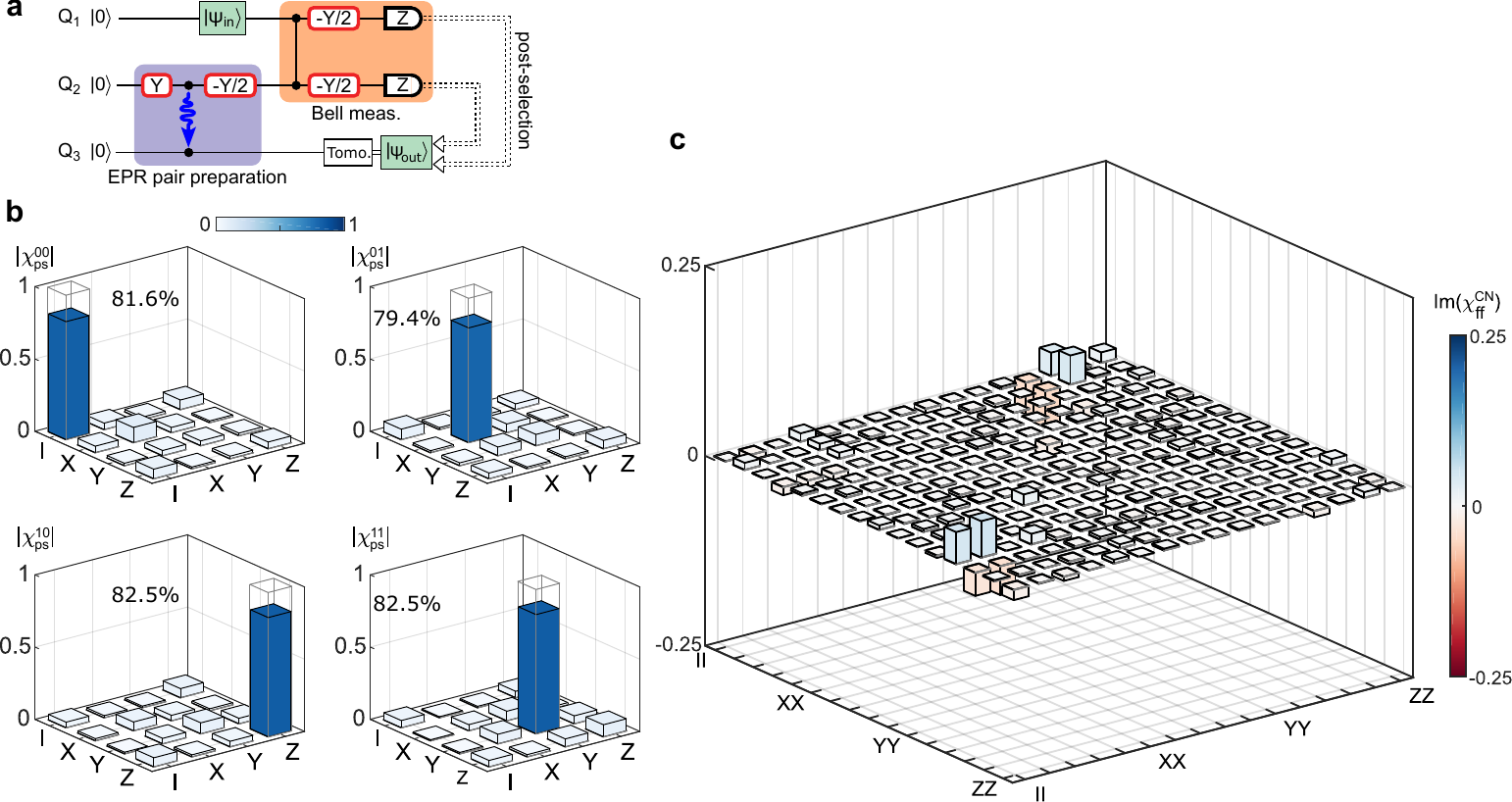}
  \caption{\label{teleport} {\bf Supplementary data for teleportation.}
  {\bf a}, {\bf b}, The protocol and process matrix  $\chi_{{\rm ps}}^{ij}$ for quantum state teleportation from $Q_1$ to $Q_3$ with post-selection.
  {\bf c}, The imaginary part of the process matrix  $\chi_{{\rm ff}}^{CN}$ for deterministic teleportation of the CNOT gate.
  The solid bars and grey frames are the measured and ideal values respectively.
  }
\end{figure}
By sacrificing the deterministic feature, one could avoid the 1.3~$\mu$s latency in feed-forward and instead post-select the $Q_3$ outcome upon the classical measurement results, see the protocol in Fig.~\ref{teleport}{\bf a}.
Conditioned on the $Q_1$-$Q_2$ joint readout result of $ij=00$, 01, 10 or 11, we reconstruct the $Q_3$ quantum state $|\psi_{\rm out}\rangle$, and calculate the corresponding process matrices $\chi_{{\rm ps}}^{ij}$, as shown in Fig.~\ref{teleport}{\bf b}.
We find $\chi_{{\rm ps}}^{ij}$ has a process fidelity of $81.6\pm1.2\%$~(00), $79.4\pm 1.4\%$~(01), $82.5\pm 1.2\%$~(10) and $82.5\pm1.3\%$~(11) respectively, with an average process fidelity of 81.5\%.

In Fig.~4{\bf d} in the main text, the real part of the teleported CNOT gate process matrix $\textrm{Re}(\chi_{\textrm{ff}}^{CN})$ is displayed. The imaginary part $\textrm{Im}(\chi_{\textrm{ff}}^{CN})$ is shown in Fig.~\ref{teleport}{\bf c}, which should be zero ideally. 

\clearpage

\bibliographystyle{naturemag}
\bibliography{bibliography}